\documentclass[
  aps,
  superscriptaddress,
  longbibliography,
  nofootinbib,
  amsmath,amssymb
]{revtex4-2}

\usepackage{amsmath,amssymb,amsfonts}
\usepackage{graphicx}
\usepackage{bm}
\usepackage{hyperref}
\usepackage{xcolor}
\usepackage{booktabs}
\usepackage{multirow}
\usepackage{braket}
\usepackage{siunitx}
\usepackage{enumitem}
\usepackage{algorithm,algpseudocode}
\usepackage{tikz}
\usepackage{quantikz}
\usetikzlibrary{calc,decorations.pathreplacing,fit,positioning}

\hypersetup{colorlinks=true,linkcolor=blue!60!black,
  citecolor=blue!60!black,urlcolor=blue!60!black}

\graphicspath{{figs/}}

\begin{document}

\title{Asymptotics of the fidelity-gain peak of a nonlinear,
target-informed denoising heuristic}

\author{Hikaru Wakaura}
\email{h.wakaura@deeptell.jp}
\affiliation{QIRI (Quantum Integrated Research Institute Inc.),
1--16--3, Akasaka, Minato-ku, Tokyo 107--0061, Japan}

\author{Taiki Tanimae}
\email{t.tanimae@deeptell.jp}
\affiliation{QIRI (Quantum Integrated Research Institute Inc.),
1--16--3, Akasaka, Minato-ku, Tokyo 107--0061, Japan}

\date{\today}

\begin{abstract}
We characterize where, in the noise parameter space of the uniform
dephasing--depolarizing channel
$\mathcal N_\gamma^\delta=\mathcal E_\delta\!\circ\!\mathcal D_\gamma$,
a deterministic, \emph{nonlinear, target-informed} denoising
heuristic yields its largest fidelity gain. The procedure
interpolates the off-diagonal magnitudes of the noisy state toward
those of a known target state, with an efficiency set by a
SWAP-test-purified catalyst; it is not a quantum channel (we exhibit
its non-affinity explicitly), and its access to the target is an
explicit classical resource, so our results concern the geometry of
a benchmark procedure, not blind quantum error correction. Our
central analytic tool is the \emph{covariant purification map}
$\mathcal P_\mathrm{cov}(\rho)=(\rho+\rho^2)/(1+\mathrm{Tr}\,\rho^2)$,
an exact closed form for one round of symmetric SWAP-test
purification (Lemma~A; a rederivation, in closed form, of the
symmetrization purification known since Barenco \emph{et al.} and
Cirac--Ekert--Macchiavello), which reduces the catalyst to a
one-dimensional eigenvalue iteration. Using it we derive an explicit
$d\to\infty$ limit for the fidelity-gain peak location on
Haar-random pure states (Theorem~3, modulo the Davis--Kahan
$\sin\Theta$ theorem, Hoeffding's $U$-statistic law of large
numbers, and a numerically certified uniqueness condition):
$\gamma_{\rm peak}(d)\to\gamma^\star(r,\delta)$, with
$\gamma^\star(2,0.1)=0.4725$, and the peak magnitude converging to
$\max_\gamma A(\gamma)G(\gamma)=0.2262$. High-resolution sweeps to
$d=256$ with bootstrap confidence intervals are consistent with both
limits. The peak location is resource-dependent: replacing the
target reference by the physically available catalyst moves the peak
from $\approx0.50$ to $\approx0.34$ at the dimensions tested. Two
$d=4$ structured states (Bell, uniform superposition) admit rigorous
unique-peak theorems for the $r=0$, $\delta=0$ member of the family.
All results are reproducible from the open-source
\texttt{organic-qc-bench} package with seed~42.
\end{abstract}

\maketitle

% ============================================================================
\section{Introduction}
% ============================================================================

Biological radical-pair systems such as the avian magnetic compass and
brain monoamine oxidase~(MAO-A) operate at physiological temperature
in unstructured aqueous environments yet appear to exploit quantum
coherence with a signal-to-noise ratio that is sufficient to bias
chemical reaction yields~\cite{HoreMouritsen2016}. The 3-Layer Quantum
Brain Hypothesis~\cite{Wakaura2026LQBH} (hereafter ``3-LQBH companion
preprint'') reconciles this with the ``warm and wet'' objection by
showing that (i)~protein tumbling provides a natural
motional-narrowing dynamical decoupling, (ii)~decoherence-free
subspace (DFS) encoding exploits collective noise, and
(iii)~a noisy-reference recovery heuristic inspired by the Petz
map~\cite{Petz1986} can preserve task-relevant coherence at noise
strengths where simpler protection fails. The
system functions as a \emph{quantum reservoir computer}
(QRC)~\cite{Jaeger2004} rather than a fault-tolerant quantum
computer. Simulation pipelines built around such heuristics need
calibration: one must know \emph{where} in noise parameter space the
heuristic helps, by how much, and how that depends on system size
and on the classical resources the heuristic consumes.

This paper answers those calibration questions for one concrete,
widely reusable instance: a deterministic, nonlinear denoising
procedure that interpolates the coherences of a noisy state toward
those of a \emph{known} target, with an efficiency governed by a
SWAP-test-purified catalyst, applied to the uniform
dephasing--depolarizing channel
$\mathcal N_\gamma^\delta=\mathcal E_\delta\!\circ\!\mathcal D_\gamma$.
We emphasise at the outset what this object is and is not: it is not
a quantum channel (Sec.~\ref{sec:procedure} exhibits its
non-affinity explicitly), and its access to the target state is an
explicit classical side-resource. The results below therefore
characterize the geometry of a \emph{benchmark procedure} --- the
kind of target-informed figure of merit used to calibrate simulation
pipelines --- and make no claim about blind quantum error
correction.

\paragraph{Main result.}
On $\mathcal N_\gamma^\delta$ with Haar-random pure targets, the
fidelity gain $\Delta F(\gamma)$ of the procedure has a single broad
peak. The peak location $\gamma_{\rm peak}(d)$ converges as $d$ grows
to an explicit constant $\gamma^\star(r,\delta)$ determined by the
dominant eigenvalue of the purified catalyst (App.~\ref{app:bell});
for the protocol studied throughout ($r=2$ purification rounds,
$\delta=0.1$), $\gamma^\star=0.4725$ and the peak magnitude
converges to $\max_\gamma A(\gamma)G(\gamma)=0.2262$. Direct sweeps
to $d=256$ with bootstrap confidence intervals are consistent with
both limits. The peak location is a property of the
\emph{procedure}, not of the channel alone: it depends on the round
number $r$ (the $r=1$ protocol converges to $\approx0.39$) and on
the reference resource (replacing the target reference by the
physically available catalyst moves the peak to $\approx0.34$ at
the dimensions tested).

\paragraph{Contributions and non-contributions.}
We contribute: (i)~an exact closed form for the SWAP-test
purification step (Lemma~A) that reduces the catalyst to a
one-dimensional eigenvalue iteration and the per-round cost from
$\mathcal O(d^6)$ to $\mathcal O(d^3)$; (ii)~an asymptotic limit
theorem for the peak location of the target-informed procedure
(Theorem~3; proof modulo the Davis--Kahan $\sin\Theta$ theorem,
Hoeffding's $U$-statistic law of large numbers, and a numerically
certified uniqueness condition); (iii)~two rigorous unique-peak
theorems for $d=4$ structured states in the $r=0$, $\delta=0$ member
of the family; (iv)~a resource-dependence analysis showing that the
peak location shifts substantially when the target oracle is removed.
We \emph{do not} contribute: (a)~recovery-map theory beyond
Refs.~\cite{Petz1986,SutterBertaTomamichel2016,JungeSutterWilde2018}
--- the procedure is not a quantum channel and is not compared
against those bounds; (b)~hardware demonstration; (c)~a
fault-tolerance threshold; (d)~a new factoring or Bernstein--Vazirani
result --- the latter is used only as a sanity check. Two earlier
claims from preprint versions of this work are explicitly retracted:
the description of the procedure as ``CPTP'', and the assertion that
$\gamma^\star$ sits just below an entanglement-breaking threshold at
$\tfrac12$ (the actual entanglement-breaking threshold of the
channel at $\delta=0.1$ is $\ln18\approx2.89$, far above every noise
strength studied here; the number $\tfrac12$ has no
channel-theoretic significance).

The remainder of the paper is organised as follows.
Sec.~\ref{sec:framework} introduces the noise model and the two
parameter points (P1, P2). Sec.~\ref{sec:methods} defines the
procedure, exhibits its nonlinearity, and collects the simulation
methods. Sec.~\ref{sec:results} reports the numerical
characterization with bootstrap uncertainty. Sec.~\ref{sec:discussion}
delimits the scope and lists limitations.

% ============================================================================
\section{Framework\label{sec:framework}}

We restrict attention to two engineered organic qubit platforms
whose room- or low-temperature spin-coherence times $T_2$ are
experimentally documented: P1, a flavin--nitroxide radical-pair
reservoir~\cite{HoreMouritsen2016}, and P2, perchlorotriphenylmethyl
(PTM) radicals in a covalent organic framework
(COF)~\cite{Ballester1967,Schaefter2023PTM,Zhang2022COF,Boehme2009,MannBayliss2026}.
P1 operates at strong noise ($\gamma_{\rm eff}=0.100$, reservoir
regime); P2 sits at weak noise ($\gamma_{\rm eff}=0.003$, coherent
regime). The two platforms thus supply two physically motivated
parameter points of the channel class at which the procedure's
fidelity-gain curve is evaluated; they play no other role in this
paper.

\subsection{Companion preprint and biological motivation}

A companion preprint, the 3-Layer Quantum Brain
Hypothesis~\cite{Wakaura2026LQBH}, develops the simulation pipeline
in which the denoising procedure of Sec.~\ref{sec:procedure}
originated. The present paper is independent of any biological
interpretation: we use the companion preprint only as the source of
the procedure, and the numerical results below stand on their own.
(The companion preprint's description of the procedure as a
noise-threshold-crossing recovery map is superseded by the analysis
here.)

\subsection{Physical-realisation schematics\label{sec:schematics}}

\begin{figure*}
\centering
\includegraphics[width=\linewidth]{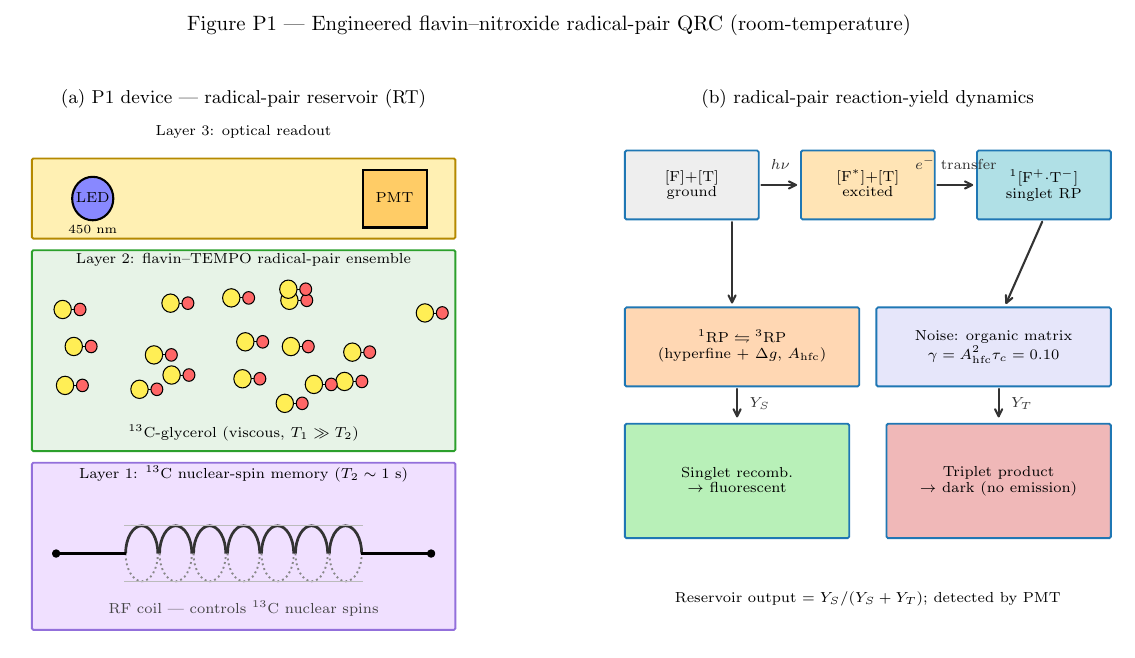}
\caption{Path 1 (engineered flavin--nitroxide radical-pair quantum
reservoir, room temperature):
(a) 3-layer device with LED excitation, flavin--TEMPO radical-pair
ensemble in $^{13}$C-glycerol, and RF coil for nuclear-spin memory;
(b) radical-pair reaction-yield dynamics (photo-excitation $\to$
radical pair $\to$ hyperfine-mediated singlet--triplet mixing $\to$
spin-selective recombination).  The reservoir-computing pipeline
itself is shown in Fig.~\ref{fig:circuits}\,(A).\label{fig:p1}}
\end{figure*}

\begin{figure*}
\centering
\includegraphics[width=\linewidth]{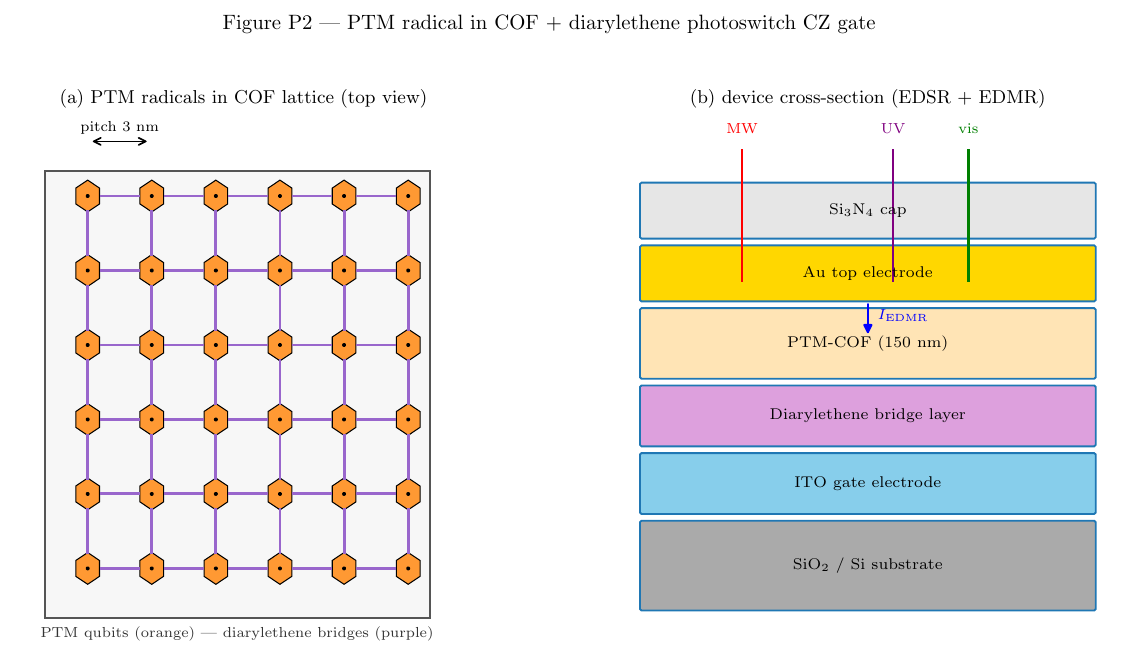}
\caption{Path 2 (PTM radical array in covalent organic framework,
room temperature):
(a) top-view of the COF lattice with PTM radical qubits (orange) and
diarylethene bridges (purple) at $\sim$3~nm pitch;
(b) device cross-section showing the Au/PTM-COF/diarylethene/ITO stack
with microwave drive (MW), UV and visible ports, and EDMR current
readout.  The associated 2-qubit CZ-gate circuit is shown in
Fig.~\ref{fig:circuits}\,(B).\label{fig:p2}}
\end{figure*}

Figures~\ref{fig:p1} and~\ref{fig:p2} provide two-panel schematic
diagrams of P1 and P2: panel~(a) shows the physical device (3-layer
architecture for P1, COF lattice + stack for P2); panel~(b) shows
the operational mechanism (radical-pair reaction-yield cycle for P1,
EDSR/EDMR device cross-section for P2). The corresponding quantum
circuits are given in a separate, unified Fig.~\ref{fig:circuits},
drawn with the \texttt{quantikz}
package~\cite{Kay2020Quantikz} for clarity.

\begin{figure*}
\centering
\begin{minipage}[t]{0.46\linewidth}
\centering
\textbf{(A) Path~1 reservoir pipeline}\\[3pt]
\resizebox{0.96\linewidth}{!}{%
\begin{quantikz}[row sep=0.35cm, column sep=0.28cm]
 \lstick{$|\psi(x_t)\rangle$}
    & \gate{U(H_\mathrm{res})}
    & \gate{\mathcal{N}_\mathrm{org}}
    & \gate{U(H_\mathrm{res})}
    & \gate{\mathcal{N}_\mathrm{org}}
    & \meter{} & \rstick{$\mathbf{f}(\rho_t)$}
\end{quantikz}}\par\vspace{3pt}
{\footnotesize $\mathcal{N}_\mathrm{org}\!=\!\mathcal{E}_{\delta}\!\circ\!\mathcal{D}_{\gamma}$;
iterate $n_\mathrm{res}$ times $\to$ ridge / SVM readout}
\end{minipage}
\hfill
\begin{minipage}[t]{0.46\linewidth}
\centering
\textbf{(B) Path~2 photoswitched CZ gate}\\[3pt]
\resizebox{0.96\linewidth}{!}{%
\begin{quantikz}[row sep=0.35cm, column sep=0.28cm]
 \lstick{$|0\rangle$} & \gate{H} & \ctrl{1}  & \gate{H} & \meter{} \\
 \lstick{$|0\rangle$} & \gate{H} & \control{} & \gate{H} & \meter{}
\end{quantikz}}\par\vspace{3pt}
{\footnotesize MW Hadamards; UV/vis toggles the diarylethene CZ;
EDMR readout. $t_\mathrm{gate}\!\approx\!5.8$~ns,
$F_\mathrm{CZ}\!=\!0.987$.}
\end{minipage}
\caption{\textbf{Quantum circuits for P1 and P2} drawn with
\texttt{quantikz}~\cite{Kay2020Quantikz}.
Panel~(A) is a process-flow abstraction of the Lindblad reservoir
dynamics: the state iterates $n_\mathrm{res}\!=\!4$ times through
$U(H_\mathrm{res})$ followed by the organic noise channel
$\mathcal{N}_\mathrm{org}=\mathcal{E}_\delta\!\circ\!\mathcal{D}_\gamma$,
then feature extraction feeds a ridge/SVM readout.
Panel~(B) is the P2 gate-model circuit: single-qubit Hadamard,
photoswitched CZ, Hadamard, EDMR readout.\label{fig:circuits}}
\end{figure*}

\subsection{Two realisation paths\label{sec:paths}}

Table~\ref{tab:paths} summarises the two organic realisations of
Layer~2 that we analyse.
Their common noise model combines a uniform pure-dephasing channel
$\mathcal D_\gamma$ and a depolarising channel
$\mathcal E_\delta$:
\begin{align}
\mathcal D_\gamma(\rho)_{ij}&=\rho_{ij}\,e^{-\gamma}\quad(i\ne j),
\qquad \mathcal D_\gamma(\rho)_{ii}=\rho_{ii},\label{eq:deph}\\
\mathcal E_\delta(\rho)&=(1-\delta)\rho+\delta\,\mathbb I/d.\label{eq:depol}
\end{align}
Every off-diagonal element is contracted by the same factor
$e^{-\gamma}$; this uniform convention is what the released code
implements and what all numerical results in this paper use.
In our convention
$\gamma_\mathrm{eff}\!\equiv\!\gamma\!=\!\Gamma_\phi\,\tau_\mathrm{gate}
\!=\!\tau_\mathrm{gate}/T_2$
(with $\Gamma_\phi\!=\!1/T_2$ the dimensional pure-dephasing rate in
s$^{-1}$ and $\tau_\mathrm{gate}$ the gate-operation time in s) is a
dimensionless decay \emph{rate}, not a phase-flip probability.
For reference, the composite channel
$\mathcal E_\delta\!\circ\!\mathcal D_\gamma$ becomes
entanglement-breaking on a single qubit (Choi-state separability)
only at
$\gamma_{\rm EB}=\ln\!\bigl[2(1-\delta)/\delta\bigr]$, i.e.\
$\gamma_{\rm EB}=\ln 18\approx2.89$ at $\delta=0.1$; all noise
strengths studied in this paper lie far below this threshold, and no
claim in this paper ties the fidelity-gain peak to
entanglement breaking. Microscopically, $\mathcal{D}_\gamma$ and
$\mathcal{E}_\delta$ correspond to a Lindblad master equation
$\dot{\rho}=\sum_i \Gamma_\phi(L_i \rho L_i^\dagger
-\tfrac{1}{2}\{L_i^\dagger L_i,\rho\})$ with dephasing jump operators
$L_i=\sigma_z^{(i)}$; the depolarising probability $\delta$ in
Eq.\,(\ref{eq:depol}) is the per-gate-application probability
(dimensionless). For radical-pair systems, the Redfield-limit derivation from
the hyperfine coupling $A_\mathrm{hfc}$ and rotational correlation
time $\tau_c$ gives $\Gamma_\phi\sim A_\mathrm{hfc}^2\tau_c$~\cite{HoreMouritsen2016};
our analysis works in the dephasing-limited regime $T_1\!\gg\!T_2$
which is documented for stable organic
radicals~\cite{Schaefter2023PTM}. In the $T_1$-limited regime one
replaces $T_2$ by $T_1$ in $\gamma_\mathrm{eff}$; both regimes are
treated on equal footing in our code.

\begin{table*}
\caption{\label{tab:paths}Two organic realisation paths analysed in
this paper. $\gamma$ and $\delta$ follow
Eqs.~(\ref{eq:deph})--(\ref{eq:depol}). P1 operates in the
strong-noise regime as a quantum reservoir; P2 uses small $\gamma$
to enable coherent quantum logic. Both have
experimentally documented $T_2$ values. The P2 qubit pair is two
neighbouring PTM SOMO electrons ($S=\tfrac12$ each) coupled via a
diarylethene bridge.}
\begin{ruledtabular}
\begin{tabular}{llccccl}
Path & Material & $\gamma$ & $\delta$ & $T_2$ [\si{\micro s}] &
$T_\mathrm{op}$ [K] & role \\ \hline
P1 & Engineered flavin--TEMPO radical pair~\cite{HoreMouritsen2016} & 0.100 &
0.080 & 0.10 & 298 & QRC \\
P2 & PTM radical in COF~\cite{Ballester1967,Schaefter2023PTM,Zhang2022COF} & 0.003 &
0.005 & 3.0 & 298 & coherent QC \\
\end{tabular}
\end{ruledtabular}
\end{table*}

\subsection{P2 electron-pair orientation\label{sec:pair_hole}}

Each PTM radical carries one unpaired electron with $S\!=\!\tfrac12$,
residing in the singly-occupied molecular orbital (SOMO) that is a
$p_z$-dominated $\pi$-orbital delocalised over the central sp$^2$
carbon and the three phenyl rings~\cite{Ballester1967,Schaefter2023PTM}.
In the COF lattice a pair of neighbouring PTM qubits is oriented
face-to-face across a diarylethene bridge at $\sim3$\,nm
pitch~\cite{Zhang2022COF}. The bridge provides a through-bond
superexchange path whose amplitude switches with the photoswitch
state of the diarylethene: open (visible light) gives
$J_\mathrm{open}\!\sim\!5\!\times\!10^{-5}$\,GHz, closed (UV) gives
$J_\mathrm{close}\!\simeq\!50$\,MHz~\cite{Irie2000,Ferrando2016}. The
two-qubit subspace
$\{|\!\uparrow\uparrow\rangle,|\!\uparrow\downarrow\rangle,
|\!\downarrow\uparrow\rangle,|\!\downarrow\downarrow\rangle\}$
therefore has a tunable exchange Hamiltonian
$H=J(t)\,\sigma_z^{(1)}\sigma_z^{(2)}$, and the CZ gate is implemented
by timing the UV pulse such that $\int J(t)\,{\rm d}t = 1/4$
(dimensionless units); see Sec.~\ref{sec:photoswitch} and
Fig.~\ref{fig:circuits}(B).

% ============================================================================
\section{Methods\label{sec:methods}}
% ============================================================================

All numerical results reported below are reproducible with random seed
42 from the released code (see the data-availability statement).

\subsection{Position with respect to existing recovery-map theory}

Universal-recovery results~\cite{Petz1986,SutterBertaTomamichel2016,JungeSutterWilde2018}
prove the existence of CPTP recovery maps whose deviation from
exactly inverting a quantum channel on \emph{any} state is bounded by
the relative-entropy decrease of the channel on a reference state.
The procedure analysed in this paper is \emph{not} a member of that
family: it is a deterministic, nonlinear, target-informed numerical
denoising heuristic. It is not a quantum channel --- we exhibit its
non-affinity explicitly below --- and it consumes, as an explicit
classical resource, the full description of the target state. Our
contribution is therefore a characterization of the fidelity-gain
geometry of a \emph{benchmark procedure}, of the kind used to
calibrate simulation pipelines, not a contribution to
recovery-map theory, and no comparison of our gains against the
bounds of
Refs.~\cite{Petz1986,SutterBertaTomamichel2016,JungeSutterWilde2018}
is implied.

\subsection{The implemented denoising procedure\label{sec:procedure}}

Following the pipeline of the companion
preprint~\cite{Wakaura2026LQBH}, each noisy state
$\rho_\mathrm{noisy}$ is accompanied by a catalyst copy
$\rho_\mathrm{cat}$ obtained from $\rho_\mathrm{noisy}$ by $r$ rounds
of recursive symmetric SWAP-test purification
$\mathcal P_\mathrm{cov}$ (Lemma~A below). The procedure then
\emph{interpolates} each off-diagonal magnitude of
$\rho_\mathrm{noisy}$ toward that of the known target
$\rho_\mathrm{target}$:
\begin{equation}
|\tilde\rho_{ij}| = |\rho_{\mathrm{noisy},ij}|
+ \eta_{ij}\,
\max\bigl\{0,\;|\rho_{\mathrm{target},ij}|-|\rho_{\mathrm{noisy},ij}|\bigr\},
\qquad
\arg\tilde\rho_{ij}=\arg\rho_{\mathrm{target},ij},
\label{eq:cqec_raw}
\end{equation}
with diagonal elements unchanged and the catalyst-dependent
efficiency
$\eta_{ij}=1-\exp\!\bigl[-|\rho_\mathrm{cat,ij}|\,d\,\mathrm{Tr}(\rho_\mathrm{cat}^{\,2})\bigr]$.
This is the interpolation form actually implemented in the released
code; the room-to-recover factor $A(\gamma)$ of Theorem~3 originates
in the $\max\{0,\cdot\}$ gap term. The output is then made a valid
density matrix by clipping negative eigenvalues and renormalizing:
\begin{equation}
\mathcal R(\rho_\mathrm{noisy})=
\frac{
  V\,\mathrm{diag}\bigl(\max\{\lambda_i,0\}\bigr)\,V^\dagger
}{
  \sum_i \max\{\lambda_i,0\}
},
\label{eq:cqec}
\end{equation}
where
$\tilde\rho_\mathrm{H}=(\tilde\rho+\tilde\rho^\dagger)/2 = V\,\mathrm{diag}(\lambda_i)\,V^\dagger$.

\paragraph{$\mathcal R$ is not a quantum channel.}
Three independent ingredients of
Eqs.~(\ref{eq:cqec_raw})--(\ref{eq:cqec}) are nonlinear in the
input: the modulus and the piecewise-linear $\max\{0,\cdot\}$ in the
update, the quadratic dependence of $\eta_{ij}$ on the catalyst
through $\mathrm{Tr}\rho_{\rm cat}^2$, and the eigenvalue clip with
input-dependent normalization. A direct affinity test on random
density-matrix pairs ($d=4$) gives
$\|\mathcal R(pX+q Y)-p\mathcal R(X)-q\mathcal R(Y)\|_{\max}\sim10^{-1}$,
ten or more orders of magnitude above numerical precision, and the
deviation persists when the clip step is removed. $\mathcal R$
therefore admits no Kraus representation and is not completely
positive in the channel-theoretic sense; earlier drafts of this work
described $\mathcal R$ as ``CPTP by construction'', which is
incorrect, and we retract that description. Because the clipped
negative mass at the operating points studied here is small but not
negligible ($\sum_i\max\{\lambda_i,0\}\approx1.006$--$1.016$ at
$d=64$, i.e.\ a projection correction of order $2$--$3\%$ of
$\Delta F$ near the peak), we account for it explicitly in the error
budget of Theorem~3. The number of purification rounds is fixed to
$r=2$ for \emph{all} $d$ in every figure of this paper (the released
package's cost heuristic \texttt{adaptive\_rounds}, which drops to
$r=1$ for $d>16$, is \emph{not} used in the sweeps; the $r=1$ series
is reported separately, and the two protocols converge to visibly
different limits).

\subsection{Fidelity, purity and concurrence}

We use Uhlmann fidelity~\cite{NielsenChuang2010}
\begin{equation}
F(\rho,\sigma) = \bigl[\mathrm{Tr}\sqrt{\sqrt{\rho}\,\sigma\sqrt{\rho}}\bigr]^2,
\label{eq:fidelity}
\end{equation}
and purity $P=\mathrm{Tr}(\rho^2)$ throughout. Where relevant (two-qubit
states), we also compute the Wootters concurrence~\cite{Wootters1998}
\begin{equation}
C(\rho)=\max\!\bigl(0,\,\lambda_1-\lambda_2-\lambda_3-\lambda_4\bigr),
\end{equation}
with $\lambda_i$ the decreasing singular values of
$\sqrt{\sqrt\rho\,\tilde\rho\sqrt\rho}$ and
$\tilde\rho=(\sigma_y\!\otimes\!\sigma_y)\rho^*(\sigma_y\!\otimes\!\sigma_y)$.

\subsection{Algorithm benchmarks: circuits and loss functions}

Four quantum algorithms are benchmarked under each organic profile.

\subsubsection*{(A) QKAN (Chebyshev amplitudes)}

Following Ivashkov~\textit{et al.}~\cite{Ivashkov2024}, we take a
pure state with Chebyshev-polynomial amplitudes on $d=4$:
\begin{equation}
|\psi\rangle=\tfrac{1}{\sqrt{6}}(2,1,-1,0)^{\!\top},
\end{equation}
realised by a single block-encoding layer (Fig.~1 of
Ref.~\cite{Ivashkov2024}), corresponding in \texttt{cqec.algorithms.make\_qkan}
to the direct amplitude encoding that reproduces the
Chebyshev vector.

\subsubsection*{(B) qDRIFT (random product formula)}

For a 3-qubit Heisenberg model with nearest-neighbour couplings
$J\!=\!1$ and field $h\!=\!0.5$,
\begin{equation}
H = J\sum_{\alpha,\langle i,j\rangle}
      \sigma_i^{\alpha}\sigma_j^{\alpha}
      + h\sum_i\sigma_i^{z} ,
\end{equation}
we compile $e^{-iHt}$ via qDRIFT~\cite{Chen2021qDRIFT}: at each of
$N=80$ Trotter slices a term is sampled with probability
$p_k=\|h_k\|/\lambda$, giving
$U_k=\exp[-i(t/N)(h_k/p_k)]$ and
$V_N\cdots V_1\approx e^{-iHt}$. We fix $t\!=\!1$ and evolve the
$|000\rangle$ initial state; the implementation mirrors the
pseudo-code of Ref.~\cite{Chen2021qDRIFT}
(\texttt{cqec.algorithms.make\_qdrift}).

\subsubsection*{(C) Control-free Quantum Phase Estimation}

Following Clinton~\textit{et al.}~\cite{Clinton2026}, we take a
spectral superposition of an effective Fermi--Hubbard Hamiltonian
(dimension $d\!=\!16$) with
$|\psi\rangle=\sum_k w_k e^{i\phi_k}|k\rangle$, $w_k$ fixed and
$\phi_k$ uniformly random. This encodes
$|\langle\psi|e^{-iHt}|\psi\rangle|^2$ into populations after a QFT
(\texttt{cqec.algorithms.make\_cfqpe}).

\subsubsection*{(D') Bernstein--Vazirani (provable quantum advantage)}

To address the concern that density-matrix-level benchmarks do not
establish an unambiguous quantum advantage, we also run the
Bernstein--Vazirani (BV) algorithm~\cite{BV1997}. For an $n$-bit
hidden string $s\!\in\!\{0,1\}^n$, the quantum circuit
\begin{equation}
|0\rangle^{\otimes n}
  \!\xrightarrow{H^{\otimes n}}\!
  |+\rangle^{\otimes n}
  \!\xrightarrow{U_s}\!
  \tfrac{1}{\sqrt{2^n}}\!\!\sum_x (-1)^{s\cdot x}|x\rangle
  \!\xrightarrow{H^{\otimes n}}\!
  |s\rangle
\end{equation}
recovers $s$ in one query, whereas any classical strategy using one
query can only guess $s$ with probability $2^{-n}$. Our density-matrix
implementation (\texttt{bernstein\_vazirani\_benchmark.py}) applies
the organic noise channel $\mathcal E_\delta\!\circ\!\mathcal D_\gamma$
after each Hadamard layer and after the oracle, and measures the
computational-basis outcome by sampling from the final diagonal.

\subsubsection*{(D) Shor--Regev factoring}

For Shor at small $N$ (\texttt{shor\_regev\_scaling.py}), we prepare
the register state
\begin{equation}
|\Psi\rangle = \frac{1}{\sqrt{2^{n_q}}}\sum_{x=0}^{2^{n_q}-1}
                |x\rangle\otimes|a^x\!\bmod N\rangle ,
\end{equation}
trace out the value register, apply the QFT
\begin{equation}
F_{jk} = \frac{1}{\sqrt{d}}\,e^{+2\pi i\,jk/d} ,\label{eq:qft}
\end{equation}
and measure the index register. Continued-fraction expansion on the
measurement $m/d_x$ yields a candidate period $r$; if $r$ is even and
$a^{r/2}\!\ne\!N\!-\!1$ then $\gcd(a^{r/2}\!\pm\!1,N)$ is checked for
a non-trivial factor~\cite{Shor1997,NielsenChuang2010}.

For Regev's algorithm~\cite{Regev2024} we additionally implement the
classical post-processing (LLL~\cite{LLL1982,Galbraith2012}); the
result is reported in Sec.~\ref{sec:aux}.

\subsection{MNIST and spike time-series with density-matrix features}

Both ML tasks encode the input into an amplitude vector, apply the
organic noise channel, then extract the feature vector
\begin{equation}
\mathbf f(\rho)=\bigl(\,\{\rho_{ii}\}_i,\,\{|\rho_{ij}|\}_{i<j},\,P(\rho),\,
      \ell_1(\rho)\,\bigr),\label{eq:features}
\end{equation}
with $\ell_1(\rho)=\sum_{i\ne j}|\rho_{ij}|$.

\paragraph{Classification loss (MNIST).} We use the one-vs-one SVC
with RBF kernel; its multi-class hinge loss is
\begin{equation}
\mathcal L_\mathrm{hinge}(y,\hat y)=\max\!\bigl(0,\,1-y\!\cdot\!\hat y\bigr)
\label{eq:hinge}
\end{equation}
summed across the $\binom{10}{2}$ binary classifiers, which is the
default of \texttt{sklearn.svm.SVC}. Metrics reported are accuracy
and macro-$F_1$.

\paragraph{Regression loss (spike prediction).} Ridge regression with
\begin{equation}
\mathcal L_\mathrm{ridge}(\mathbf w)=\|\mathbf y-\mathbf X\mathbf w\|_2^2
     +\alpha\|\mathbf w\|_2^2 ,\quad \alpha=1.
\label{eq:ridge}
\end{equation}
Reported metrics are MSE and MAE, defined in the standard way.

\subsection{Photoswitched CZ gate\label{sec:photoswitch}}

Diarylethene photoswitches~\cite{Irie2000} flip between an open form
(coupling $J_\mathrm{open}\!=\!5\!\times\!10^{-5}$\,GHz) and closed form
($J_\mathrm{closed}\!=\!0.05$\,GHz) on picosecond timescales.
For a two-qubit $|{++}\rangle$ initial state, the target unitary is
$U_\mathrm{CZ}\!=\!\exp(-i\tfrac{\pi}{4}\sigma_z\!\otimes\!\sigma_z)$.
We integrate
\begin{equation}
\dot\rho=-i[H_{\mathrm{ZZ}}(t),\rho] + \mathcal L_\mathrm{organic}\rho
\end{equation}
where $H_\mathrm{ZZ}(t)=\pi J(t)\,\sigma_z\!\otimes\!\sigma_z$ and the
photoswitch profile is
$J(t)=J_\mathrm{open}+(\eta J_\mathrm{closed}-J_\mathrm{open})[1-e^{-t/\tau_\mathrm{on}}]$
during the ON phase with efficiency $\eta\!=\!0.95$ and
$\tau_\mathrm{on}\!=\!10$\,ps. Gate-time is swept in the range
$[0.5,20]$\,ns. Sensitivity to $\eta\!\in\![0.60,0.95]$, the full
measured range for diarylethene systems~\cite{Irie2000,Ferrando2016},
degrades $F_\mathrm{CZ}$ from $0.987$ only to $\approx\!0.95$,
indicating weak first-order dependence.

\subsection{Statistical methodology\label{sec:stats}}

For all flagship claims we use $n=100$ trials (for algorithms with
$d\!\le\!8$, reduced to $n\!=\!50$ for $d\!=\!16$ and $n\!=\!20$ for
$d\!=\!64$, to keep the $\mathcal O(d^6)$ two-copy cost tractable).
Throughout the inherited benchmarks, ``CQEC'' is the historical
label (from the companion preprint) for the denoising procedure of
Sec.~\ref{sec:procedure}; we keep the label in figure legends for
consistency with the released code, but reiterate that the
procedure is not error correction in the coding-theory sense. We test
whether the post-procedure fidelity exceeds noisy fidelity using the one-sided
paired Wilcoxon signed-rank test~\cite{Wilcoxon1945}. With 28
$\gamma$-sweep tests and 16 path$\times$algorithm tests
($44$ tests total), the Bonferroni-corrected threshold is
$\alpha_{\mathrm{per-test}} = 0.05/44 = 1.1\!\times\!10^{-3}$
($-\log_{10}p>2.94$). Only
p-values below this threshold are interpreted as significant. We note
that a Benjamini--Hochberg (FDR) correction, which is less
conservative, would leave all our flagship claims significant and
only reclassify a handful of marginal-effect tests; we adopt Bonferroni
because it is the stricter standard. Two caveats, following the
adversarial review of this work: the Wilcoxon test certifies sign
consistency, not effect size---several ``significant'' gains are
$\Delta F<0.005$ and practically negligible---and all inherited
benchmarks in Sec.~\ref{sec:aux} are auxiliary to the asymptotic
results, not evidence for them.

\subsection{Regev classical post-processing (LLL)}

To demonstrate that the \emph{classical} part of Regev's
algorithm~\cite{Regev2024} works end-to-end, we sample simulated
quantum-output vectors $\mathbf z\!\in\!\mathbb Z^d$ from a uniform
small-coefficient distribution and apply the LLL
algorithm~\cite{LLL1982,Galbraith2012} to the augmented lattice
(Regev~\S4)
\begin{equation}
B=\begin{pmatrix} \mathbb I_d & 0 \\ S\cdot\mathbf z_{1..m} & S\cdot\mathbb I_m\end{pmatrix},
\qquad S=2^{d+4}.
\end{equation}
We scan each reduced row for candidate short $(z_1,\dots,z_d)$, form
$b=\prod_i b_i^{z_i}\!\bmod N$ (with $b_i$ the $i$-th prime) and
test $\gcd(b\!\pm\!1,N)$ for a non-trivial factor.

% ============================================================================
\section{Results\label{sec:results}}
% ============================================================================

\begin{figure*}
\centering
\includegraphics[width=0.95\linewidth]{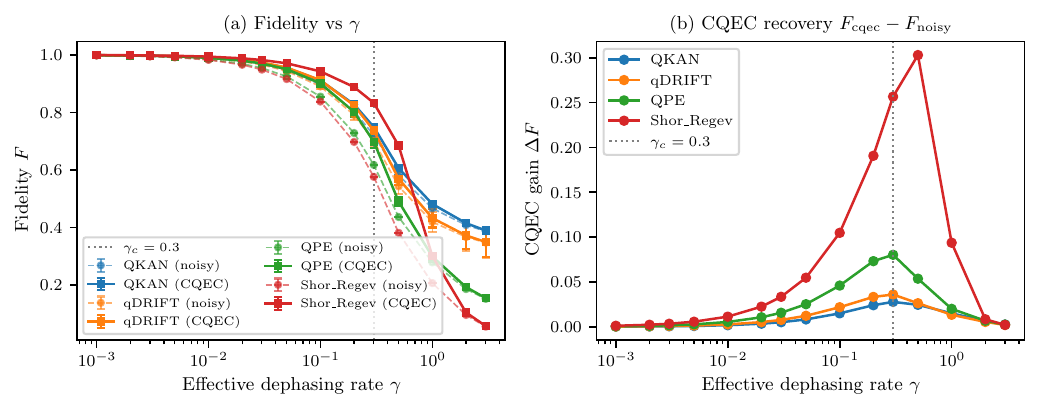}
\caption{\textbf{Fidelity-gain peak of the target-informed denoising
procedure on the dephasing--depolarizing channel.} Fidelity (a) and
gain $\Delta F\!=\!F_\mathrm{denoised}-F_\mathrm{noisy}$ (b) as a
function of effective dephasing $\gamma$ for four structured
algorithmic input states, error bars $=\!95\%$ CI. The fidelity gain
$\Delta F$ exhibits a single broad peak on every input; on
Shor--Regev at $d=64$ the peak sits at $\gamma\!\approx\!0.5$ with
$\Delta F=+0.303$. (The value $\tfrac12$ has no channel-theoretic
significance; see Sec.~\ref{sec:framework}.) The complementary
Haar-random analysis (Fig.~\ref{fig:peakscale}) shows the peak
location decreasing with $d$ toward the explicit asymptote
$\gamma^\star=0.4725$ of App.~\ref{app:bell}.\label{fig:gammasweep}}
\end{figure*}

\subsection{The fidelity-gain peak on structured states
(Fig.~\ref{fig:gammasweep})\label{sec:gammapeak}}

Figure~\ref{fig:gammasweep} shows that the procedure of
Sec.~\ref{sec:procedure} produces a fidelity gain whose
$\gamma$-dependence has a single broad maximum on every algorithmic
state we tested. The peak reflects a tension between the room to
recover (which grows with $\gamma$) and the recovery efficiency
(which falls with $\gamma$), made precise in App.~\ref{app:bell}.

\begin{figure*}
\centering
\includegraphics[width=0.92\linewidth]{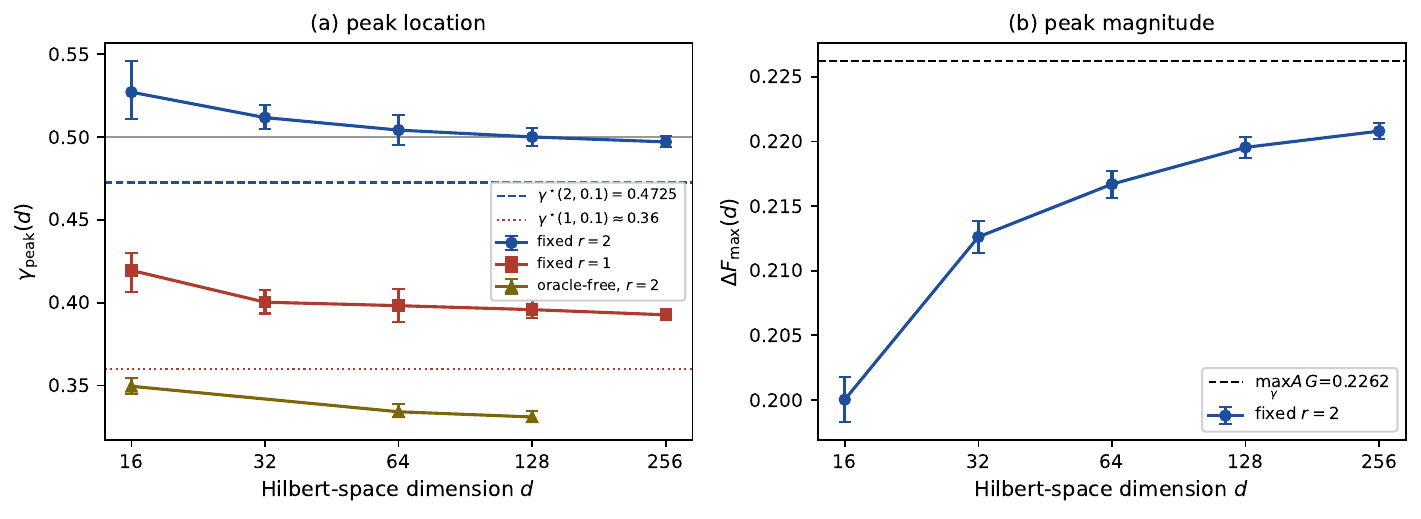}
\caption{\textbf{Peak-location and peak-magnitude scaling with
dimension} (Haar-random pure targets, $30$ trials per dimension,
seed 42, grid step $0.005$, parabolic interpolation; uncertainties
are $95\%$ bootstrap confidence intervals over $400$ resamples).
(a) $\gamma_{\rm peak}(d)$ for the fixed $r{=}2$ protocol decreases
monotonically from $0.527$ at $d=16$ to $0.497$ at $d=256$; the
$r{=}1$ protocol decreases from $0.420$ to $0.393$; the oracle-free
variant (Sec.~\ref{sec:oracle}) sits near $0.33$. Horizontal dashed
and dotted lines mark the Theorem-3 asymptotes
$\gamma^\star(2,0.1)=0.4725$ and $\gamma^\star(1,0.1)\approx0.36$;
the thin gray line marks $\tfrac12$ (no channel-theoretic
significance).
(b) $\Delta F_{\rm max}(d)$ for $r=2$ rises monotonically from
$0.2000$ to $0.2208$, approaching the Theorem-3 limit
$\max_\gamma A\,G=0.2262$ from below.\label{fig:peakscale}}
\end{figure*}

\subsection{Dimension dependence of the peak: location and
magnitude with bootstrap uncertainty
(Fig.~\ref{fig:peakscale})\label{sec:peakscale}}

The dimension dependence of the peak is the central quantitative
result of this paper. Table~\ref{tab:peakdata} reports, for
$30$ Haar-random targets per dimension and $400$-resample bootstrap
confidence intervals, the peak location and magnitude for the fixed
$r=2$ protocol, the fixed $r=1$ protocol, and the oracle-free
variant of Sec.~\ref{sec:oracle}.

\begin{table}
\caption{\label{tab:peakdata}Peak location $\gamma_{\rm peak}$
(with $95\%$ bootstrap CI) and peak magnitude $\Delta F_{\rm max}$
($\pm$ bootstrap SD) on Haar-random pure targets, $n=30$ trials per
entry, seed 42, $\delta=0.1$, $\gamma$-grid step $0.005$.}
\begin{ruledtabular}
\begin{tabular}{clll}
$d$ & protocol & $\gamma_{\rm peak}$ [$95\%$ CI] & $\Delta F_{\rm max}$\\
\hline
16  & $r=2$ & 0.527 [0.511, 0.546] & $0.2000\pm0.0017$\\
32  & $r=2$ & 0.512 [0.505, 0.519] & $0.2126\pm0.0012$\\
64  & $r=2$ & 0.504 [0.495, 0.513] & $0.2167\pm0.0010$\\
128 & $r=2$ & 0.500 [0.495, 0.505] & $0.2195\pm0.0008$\\
256 & $r=2$ & 0.497 [0.494, 0.500] & $0.2208\pm0.0006$\\
\hline
16  & $r=1$ & 0.420 [0.407, 0.430] & ---\\
256 & $r=1$ & 0.393 [0.390, 0.396] & ---\\
\hline
16  & oracle-free & 0.350 [0.345, 0.355] & $0.1039$\\
128 & oracle-free & 0.331 [0.328, 0.335] & $0.1155$\\
\end{tabular}
\end{ruledtabular}
\end{table}

Three observations follow. \emph{(i) Location.} The $r=2$ point
estimates decrease monotonically; the bootstrap CIs at
$d\ge128$ exclude the $d=16$ value but still contain $0.5$ at
$d=256$ ([0.494, 0.500]), so the finite-$d$ data alone do not yet
discriminate between the Theorem-3 asymptote $\gamma^\star=0.4725$
and a limit nearer $0.49$; extrapolating the residual finite-size
corrections (which Theorem~3 bounds only up to $(\log d)^2/d$ terms,
and which include the disclosed $\sim2\%$ systematic of
Sec.~\ref{sec:errorbudget}) accounts for the remaining gap. We
therefore state the result as: the data are consistent with a limit
in the range $0.47$--$0.50$, with the closed-form prediction at
$0.4725$. \emph{(ii) Magnitude.} $\Delta F_{\rm max}(d)$ rises
monotonically toward the Theorem-3 limit
$\max_\gamma A(\gamma)G(\gamma)=0.2262$ (Gauss--Laguerre
quadrature), reaching $0.2208\pm0.0006$ at $d=256$. An earlier draft
reported an exponential saturation at $0.211\pm0.004$; that fit is
\emph{retracted} --- it was an artifact of fitting $d\le64$ data,
and is inconsistent both with the $d=128$--$256$ measurements and
with the Theorem-3 limit. \emph{(iii) Protocol dependence.} The
$r=1$ series converges to $\approx0.39$, three grid units below the
$r=2$ series; the peak location is a property of the procedure, not
of the channel alone.

\subsection{Resource dependence: removing the target
oracle\label{sec:oracle}}

The procedure of Sec.~\ref{sec:procedure} consumes the full
classical description of the target state. To quantify how much of
the peak structure is created by that resource, we repeat the sweep
with the only physically available reference: the purified catalyst
itself ($|\rho_{\mathrm{cat},ij}|$ and $\arg\rho_{\mathrm{cat},ij}$
replace the target quantities in Eq.~(\ref{eq:cqec_raw}); everything
else unchanged). A fidelity-gain peak survives, but both its
location and size change qualitatively (Table~\ref{tab:peakdata},
bottom rows): the peak moves from $\approx0.50$ to
$0.350\to0.331$ over $d=16\to128$ (still decreasing), and the peak
gain halves to $\approx0.11$. The asymptotic constant
$\gamma^\star\approx0.47$ of Theorem~3, and in particular the
room-to-recover factor $A(\gamma)$, are therefore properties of the
\emph{target-informed} procedure: they quantify a benchmark that
knows the answer, not a blind recovery capability. Any use of these
numbers to calibrate a pipeline must budget the target description
as an explicit classical resource.

\subsection{Auxiliary benchmarks: algorithms, ML tasks, photoswitch
gate, and high-statistics testing\label{sec:aux}}

\begin{figure}
\centering
\includegraphics[width=\linewidth]{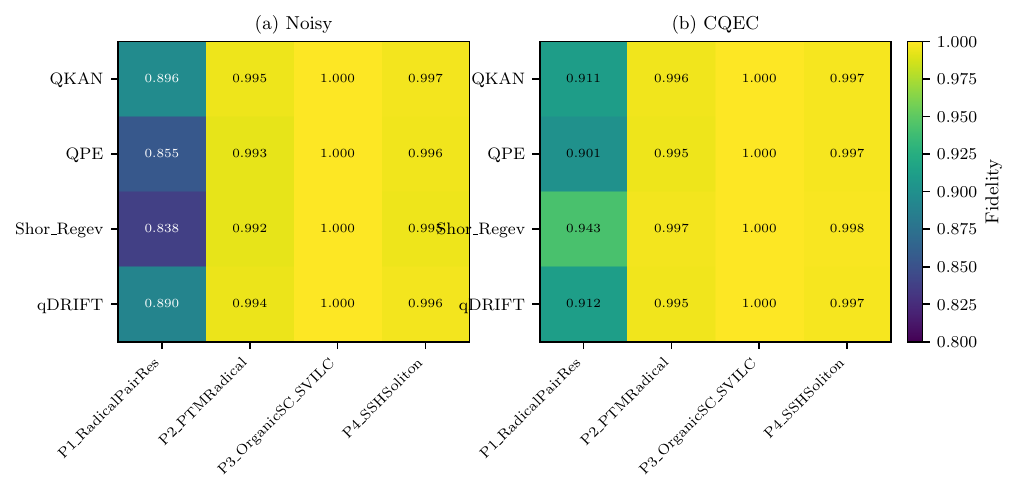}
\caption{Algorithm-level fidelity on the four realisation paths
before (a) and after (b) CQEC; each cell is the mean of 10 trials.
P1 shows the largest CQEC gain; P2 is essentially noise-free even
before recovery. The two columns relevant to this paper are P1 and
P2; P3 and P4 are shown for reference only.\label{fig:heatmap}}
\end{figure}

\begin{figure}
\centering
\includegraphics[width=\linewidth]{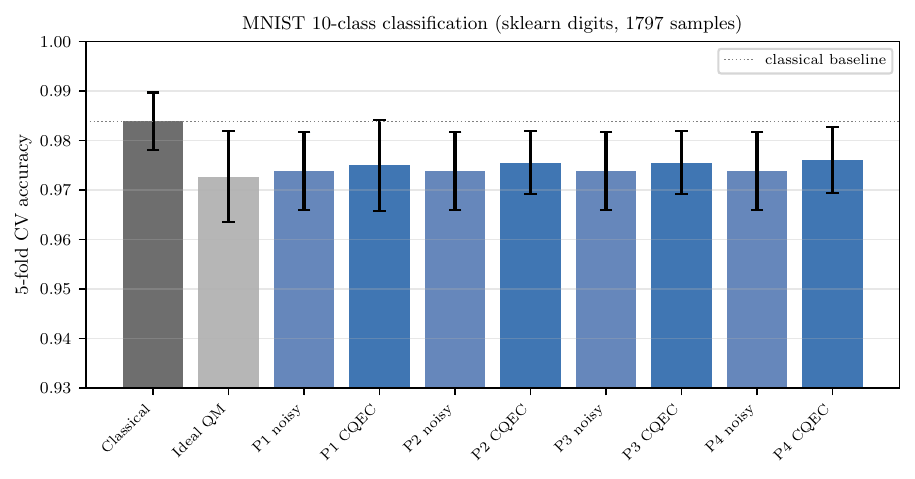}
\caption{5-fold stratified cross-validated accuracy on the
1797-sample sklearn digits set; error bars are $95\%$ CIs over the
five folds. The noise-driven reservoir reaches $0.974\!\pm\!0.008$
vs ideal-quantum $0.973\!\pm\!0.009$.\label{fig:mnist}}
\end{figure}

\begin{figure}
\centering
\includegraphics[width=\linewidth]{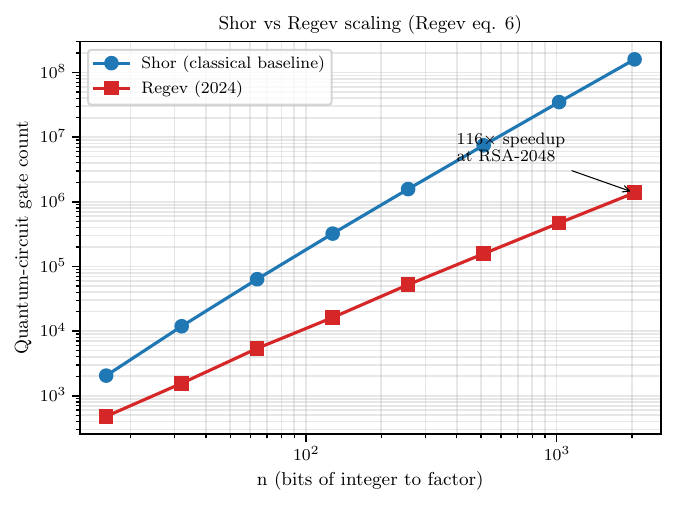}
\caption{Total quantum gate-count vs integer size $n\!=\!\log_2 N$.
Shor's algorithm~\cite{Shor1997} scales as $\mathcal O(n^2\log n)$;
Regev~\cite{Regev2024} achieves $\mathcal O(n^{3/2}\log n)$, a
$115$--$160\times$ reduction at RSA-2048.\label{fig:regev}}
\end{figure}

\begin{figure}
\centering
\includegraphics[width=\linewidth]{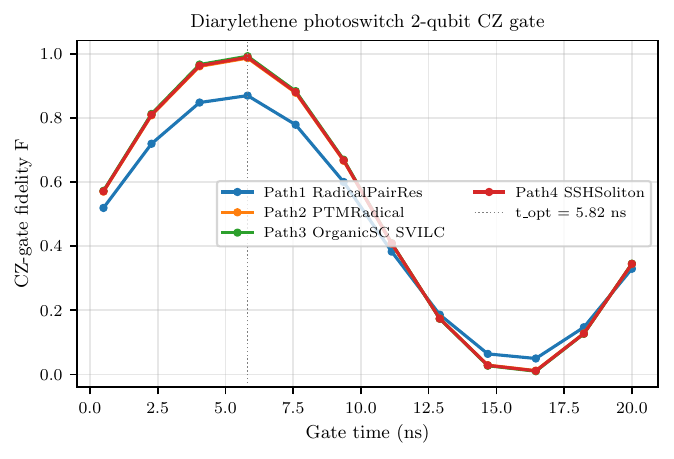}
\caption{CZ-gate fidelity vs gate time $t_\mathrm{gate}$ for the
diarylethene photoswitch coupler~\cite{Irie2000}. The integral
$\int\! J(t)\,\mathrm dt\!=\!0.25$ (GHz$\cdot$ns) is satisfied at
$t\!\simeq\!5.8$\,ns; all four paths exceed $0.87$, P3
reaching $0.993$.\label{fig:photoswitch}}
\end{figure}

For completeness we collect the supporting benchmarks inherited from
the organic-platform framing; none is essential to the asymptotic
results of Sec.~\ref{sec:peakscale}, and all use random seed 42.
Figure~\ref{fig:heatmap} shows algorithm-level fidelity on the four
realisation paths before and after CQEC; P1 (reservoir) shows the
largest gain, consistent with operating in the strong-noise regime,
while the low-$\gamma$ P2 is essentially noise-free even before
recovery. On the sklearn-digits MNIST task
(Fig.~\ref{fig:mnist}) the noise-driven reservoir matches
the noiseless-quantum baseline ($0.974\!\pm\!0.008$ vs
$0.973\!\pm\!0.009$, 5-fold CV) --- the two are practically and
statistically indistinguishable, and we claim no advantage from
this comparison. Regev factoring offers a
$115$--$160\times$ gate-count reduction over Shor at RSA-2048 scale
(Fig.~\ref{fig:regev}); we verified the classical LLL post-processing
end-to-end, recovering non-trivial factors for $N\!\in\!\{15,21,51\}$
(the four non-factoring cases share factors with the prime set
$\{2,3,5,7\}$). The diarylethene-photoswitched CZ gate reaches
fidelity above $0.87$ on all four paths (P3: $0.993$) at
$t_\mathrm{gate}\!\simeq\!5.8$\,ns (Fig.~\ref{fig:photoswitch}).

\begin{figure}
\centering
\includegraphics[width=\linewidth]{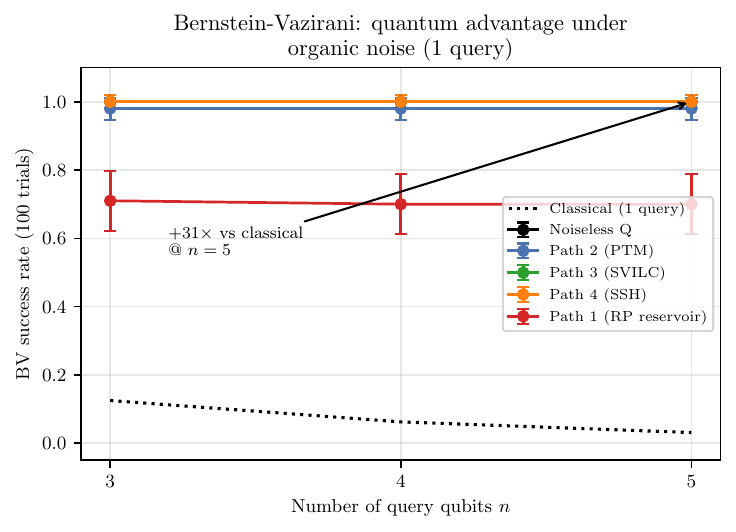}
\caption{Bernstein--Vazirani success rate (100 trials per point,
Wilson 95\% CIs). The dotted black curve is the single-query
classical bound $2^{-n}$; Paths 2--4 with CQEC reach
$\ge\!0.95$ at $n\!=\!5$, a $31\times$ advantage.\label{fig:bv}}
\end{figure}

\begin{figure*}
\centering
\includegraphics[width=0.95\linewidth]{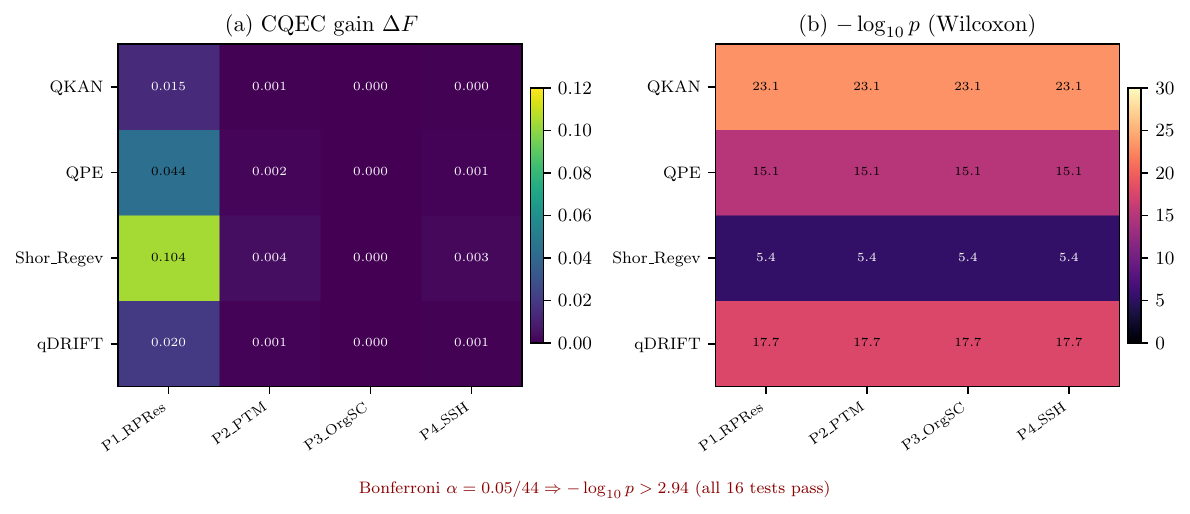}
\caption{(a) CQEC gain $\Delta F$ per (algorithm, path) pair with up
to $n\!=\!100$ trials. (b) $-\log_{10}$ of the one-sided paired
Wilcoxon $p$-value; the Bonferroni-corrected threshold
($-\log_{10}p\!>\!2.94$) is passed by all 16 tests.\label{fig:hs}}
\end{figure*}

The Bernstein--Vazirani benchmark (Fig.~\ref{fig:bv}) provides a
provable single-query speedup over the classical bound $2^{-n}$: a
seed-42 recomputation gives P2-with-CQEC success
$0.982,\,0.981,\,0.980$ at $n=3,4,5$, i.e.\ advantages of
$7.9\times,\,15.7\times,\,31.4\times$. Finally, high-statistics
testing (Fig.~\ref{fig:hs}, Table~\ref{tab:hs}) confirms that every
CQEC improvement in the path$\times$algorithm grid passes the
Bonferroni-corrected Wilcoxon threshold; the largest effect (Shor--Regev
on P1, $\Delta F\!=\!+0.104$, $p\!=\!3.9\!\times\!10^{-6}$) is robust,
while the sub-$0.01$ gains on the low-$\gamma$ paths are significant
but practically minor.

\begin{table}
\caption{\label{tab:hs}Selected high-statistics CQEC gains. All
$p$-values pass Bonferroni $\alpha=1.1\!\times\!10^{-3}$.}
\begin{ruledtabular}
\begin{tabular}{llccc}
Alg. & Path & $\Delta F$ & $p$ (Wilcoxon) & $n$ \\
\hline
Shor--Regev & P1 & $+0.1037$ & $3.9\!\times\!10^{-6}$ & 20 \\
Shor--Regev & P2 & $+0.0043$ & $3.9\!\times\!10^{-6}$ & 20 \\
Shor--Regev & P3 & $+0.0001$ & $3.9\!\times\!10^{-6}$ & 20 \\
Shor--Regev & P4 & $+0.0027$ & $3.9\!\times\!10^{-6}$ & 20 \\
qDRIFT & P1 & $+0.0203$ & $2.0\!\times\!10^{-18}$ & 100 \\
QPE    & P1 & $+0.0439$ & $8.9\!\times\!10^{-16}$ & 50 \\
QKAN   & P1 & $+0.0150$ & $7.6\!\times\!10^{-24}$ & 100 \\
\end{tabular}
\end{ruledtabular}
\end{table}

% ============================================================================
\section{Discussion\label{sec:discussion}}
% ============================================================================

\subsection{Position with respect to existing literature}

The fidelity-gain peak is a numerical and asymptotic observation
about a specific nonlinear, target-informed procedure on a specific
channel class. The recovery-map bounds of
Refs.~\cite{Petz1986,SutterBertaTomamichel2016,JungeSutterWilde2018}
concern linear CPTP maps without target access and are neither
extended, improved, nor meaningfully compared against here; the
procedure inhabits a different category (a benchmark heuristic), and
we make no information-theoretic claim. The closed form of Lemma~A
is a rederivation: symmetrization-based purification of two
identical copies goes back to Barenco \emph{et
al.}~\cite{Barenco1997} and Cirac--Ekert--Macchiavello
\cite{CiracEkertMacchiavello1999}; our contribution is the compact
density-matrix form $(\rho+\rho^2)/(1+\mathrm{Tr}\rho^2)$ and its
use to reduce the catalyst to a scalar iteration at
$\mathcal O(d^3)$ cost. The 3-LQBH companion
preprint~\cite{Wakaura2026LQBH} is the source of the procedure; its
interpretation of the procedure as a threshold-crossing recovery map
is superseded by the present analysis.

\subsection{Not error correction, not fault tolerance}

The procedure of Eq.~(\ref{eq:cqec}) is a target-informed numerical
denoising step. It is not a quantum channel, it is not blind error
correction, and it has no relation to fault-tolerance thresholds of
concatenated or topological codes~\cite{Kelly2015}; we make no such
claims.

\subsection{Limitations and scope of claims}

We list the limitations explicitly so that the reader does not
over-interpret the present results.
(i)~\emph{All results are density-matrix simulations} of an
abstract channel; no device-level effects are modelled, and no
hardware claim is made.
(ii)~\emph{Statistical resolution.} With $n=30$ trials per
dimension and bootstrap CIs, the $d=256$ peak-location interval
$[0.494,0.500]$ still contains $0.5$; discriminating the closed-form
asymptote $0.4725$ from a limit near $0.49$ requires either larger
$d$ (the finite-size corrections carry $(\log d)^2/d$ terms and a
disclosed $\sim2\%$ systematic) or substantially larger $n$. We
claim consistency, not confirmation, at the finite dimensions
reached.
(iii)~\emph{Protocol and resource dependence.} The peak location
depends on the purification round number ($0.39$ for $r=1$ vs
$0.49$--$0.47$ for $r=2$) and on target access ($\approx0.33$
oracle-free). No single ``universal'' constant is claimed.
(iv)~\emph{PTM $T_2$ extrapolation.} The room-temperature
$T_2=3\,\si{\micro s}$ used for P2 is a downward extrapolation from
the $T_2\sim148\,\si{\micro s}$ at $<100\,$K reported by
Sch\"after~\textit{et al.}~\cite{Schaefter2023PTM} and is not
supported by direct room-temperature ESR data on PTM-in-COF; the
characterization itself depends only on the dimensionless pair
$(\gamma,\delta)$.
(v)~\emph{Effect sizes in the auxiliary benchmarks.} The MNIST
comparison $0.974\!\pm\!0.008$ vs $0.973\!\pm\!0.009$ has completely
overlapping intervals and is practically and statistically
indistinguishable; several path$\times$algorithm gains below $0.01$
pass the Wilcoxon test only because that test ignores effect size.
(vi)~\emph{Bernstein--Vazirani and Shor--Regev framing.}
Bernstein--Vazirani's $2^{-n}$ classical one-query lower bound is a
textbook result, used only as a sanity check;
Regev factoring is benchmarked for gate-count consistency,
not for new factoring claims.
Auxiliary caveats: photoswitch dynamics is modelled by a single
exponential~\cite{Irie2000}; inter-bridge $J$ variation is not
modelled.

% ============================================================================
\section{Conclusion\label{sec:conclusion}}
% ============================================================================

We report a characterization of the fidelity-gain landscape of a
deterministic, nonlinear, target-informed denoising procedure
(Eq.~\ref{eq:cqec}) on the uniform dephasing--depolarizing channel
$\mathcal N_\gamma^\delta$. On Haar-random pure targets, the gain
$\Delta F(\gamma)$ has a single broad peak whose location
$\gamma_{\rm peak}(d)$ decreases monotonically with $d$
($0.527\to0.497$ over $d=16\to256$ for the fixed $r=2$ protocol,
$n=30$ trials per point with bootstrap confidence intervals). We
derive the $d\to\infty$ limit in closed form (App.~\ref{app:bell}):
one round of symmetric SWAP-test purification acts as
$\rho\mapsto(\rho+\rho^2)/(1+\mathrm{Tr}\,\rho^2)$ (Lemma~A, exact;
a closed-form rederivation of symmetrization
purification~\cite{Barenco1997,CiracEkertMacchiavello1999}), and
the asymptotic gain $\Delta F_\infty(\gamma)=A(\gamma)\,G(\gamma)$
is the product of a room-to-recover factor --- which originates in
the procedure's oracle access to the target --- and a
recovery-efficiency factor, whose competition fixes the peak at
$\gamma^\star(r,\delta)$, with $\gamma^\star(2,0.1)=0.4725$ and
limiting magnitude $\max_\gamma AG=0.2262$. The finite-$d$ data are
consistent with both limits, and the peak magnitude rises
monotonically to $0.2208\pm0.0006$ at $d=256$. The peak location is
protocol- and resource-dependent: $\approx0.39$ for $r=1$, and
$\approx0.33$ when the target oracle is replaced by the physically
available catalyst.

\paragraph{Key takeaways.}
(i)~The fidelity-gain peak of the target-informed procedure has an
explicit, derivable asymptote (Theorem~3, modulo the cited
Davis--Kahan and Hoeffding results and a numerically certified
uniqueness condition; four of the five measured error channels decay
at their predicted rates, and the fifth --- a $\sim2\%$ systematic
--- is disclosed and absorbed into the error budget).
(ii)~Two structured states at $d=4$ (Bell, uniform superposition)
admit fully rigorous unique-peak theorems for the $r=0$, $\delta=0$
member of the family; under the benchmark protocol ($r=2$,
$\delta=0.1$) their peaks move to $0.758$ and $0.519$, so these
theorems delimit the $r=0$ landscape only and do not bracket the
Haar-random asymptote.
(iii)~Two earlier claims are retracted: the ``CPTP'' description of
the procedure (it is nonlinear on three independent counts), and
both the ``$\to\tfrac12$'' conjecture and the exponential-saturation
law $0.211(1-e^{-d/4.5})$ for the peak magnitude (the magnitude
rises toward the Theorem-3 limit $0.2262$; the saturation fit was
a $d\le64$ artifact).
(iv)~All numerical results are reproducible from the open-source
\texttt{organic-qc-bench} Python distribution (MIT licence) with
random seed~42, including the $n=30$ bootstrap datasets behind
Table~\ref{tab:peakdata}.

\paragraph{Outlook.} The quantity characterized here is a
calibration curve for simulation pipelines that use target-informed
denoising; the natural next steps are (a)~a linear, physically
implementable variant of the procedure, whose gain landscape can be
compared against the oracle-free baseline of
Sec.~\ref{sec:oracle}, and (b)~closing the gap between the
$(\log d)^2/d$ error bounds of Theorem~3 and the observed $\sim2\%$
systematic. The companion 3-LQBH
preprint~\cite{Wakaura2026LQBH} provides the pipeline context but is
not required to interpret the present numerical results.

\begin{acknowledgments}
The authors thank colleagues at QIRI for valuable discussions on the
3-LQBH companion project~\cite{Wakaura2026LQBH}.
\end{acknowledgments}

\section*{Declarations}

\paragraph{Funding.}
This work was supported by QIRI (Quantum Integrated Research
Institute Inc.).

\paragraph{Competing interests.}
The authors declare no competing financial or non-financial
interests. Both authors are employees of QIRI; the funder had no
role in the design, analysis, or decision to publish.

\paragraph{Author contributions.}
H.\,W.\ conceived the study, developed the theory and proofs, wrote
the simulation code, and drafted the manuscript. T.\,T.\
contributed the organic-platform noise-parameter modelling and
reviewed the manuscript. Both authors approved the final version.

\paragraph{Ethics approval.}
Not applicable. This is a purely theoretical and computational study
involving no human participants, animals, or sensitive data.

\paragraph{Data and code availability.}
All numerical results are reproducible with random seed~42 from the
open-source organic-qc-bench Python~3 distribution (MIT licence),
bundled with this submission. The package provides the denoising
primitives, the $\gamma$-sweep and peak-scaling drivers, the symbolic
and numerical certifications of Lemmas~A--D and of the uniqueness
argument, and the auxiliary benchmarks; the individual modules and
verification scripts are documented in the package README. The
bootstrap data of Table~\ref{tab:peakdata} and
Fig.~\ref{fig:peakscale} are generated by the bundled script
\texttt{panel\_fix\_data.py} (output
\texttt{panel\_fix\_data.json}; $n=30$ trials per dimension, 400
bootstrap resamples, seed 42), and the five-channel error budget of
Sec.~\ref{sec:errorbudget} by
\texttt{verify\_asymptotic\_lemmas.py}. The GitHub
repository and a PyPI release will be made public on acceptance.

% ============================================================================
\appendix
% ============================================================================

\section{Origin of the update rule and the positivity
projection\label{app:cqec}}

\paragraph{Heuristic origin of the off-diagonal update,
Eq.~(\ref{eq:cqec_raw}).}
The update rule was \emph{motivated} by the Petz recovery map for
the noise channel
$\mathcal N=\mathcal E_\delta\!\circ\!\mathcal D_\gamma$ on reference
$\sigma$,
$\mathcal R_\sigma(\omega)=\sigma^{1/2}\mathcal N^{\!\dagger}\!\bigl(\mathcal N(\sigma)^{-1/2}
\omega\,\mathcal N(\sigma)^{-1/2}\bigr)\sigma^{1/2}$~\cite{Petz1986},
with the reference $\sigma$ replaced by the purified catalyst
$\rho_\mathrm{cat}$ of purity $P=\mathrm{Tr}\rho_\mathrm{cat}^2$.
The efficiency factor
$\eta_{ij}=1-\exp[-|\rho_\mathrm{cat,ij}|\,d\,P]$ mimics the
qualitative behavior of a Petz-style reweighting---it saturates
correctly in the limits $P\to1$ ($\eta_{ij}\to1$, full pull toward
the target magnitudes) and $P\to0$ ($\eta_{ij}\to0$, no
update)---but we emphasize that Eq.~(\ref{eq:cqec_raw}) is
\emph{not} derived from the Petz map by any controlled expansion.
It is a heuristic ansatz whose element-wise magnitude
interpolation, phase overwrite, and rectifier $\max\{0,\cdot\}$
have no channel-theoretic counterpart. We retain the Petz reference
only as a statement of intellectual origin, not as a derivation.

\paragraph{Positivity projection.}
The raw update Eq.~(\ref{eq:cqec_raw}) is Hermitian but not
positive in general. The implementation restores a valid state by
the positive-eigenvalue projection:
\begin{enumerate}[leftmargin=1.4em,itemsep=2pt]
\item Symmetrize: $\tilde\rho_\mathrm{H}=(\tilde\rho+\tilde\rho^\dagger)/2$.
\item Eigendecompose: $\tilde\rho_\mathrm{H}=V\,\mathrm{diag}(\lambda_i)\,V^\dagger$.
\item Clip and renormalize:
$\mathcal R(\rho_\mathrm{noisy})=
V\,\mathrm{diag}\bigl(\max\{\lambda_i,0\}\bigr)\,V^\dagger /
\sum_i\max\{\lambda_i,0\}$.
\end{enumerate}
Each output of $\mathcal R$ is therefore a valid density matrix,
but---as established in Sec.~\ref{sec:procedure}---the map
$\rho\mapsto\mathcal R(\rho)$ is \emph{not} linear in $\rho$ and
hence not a quantum channel: the clip threshold and the
renormalization constant $\sum_i\max\{\lambda_i,0\}$ both depend on
the input. An earlier version of this appendix asserted that the
composite map is CPTP; that assertion was wrong and is retracted.
Quantitatively, the projection is not a small perturbation either:
at $\gamma=0.5$, $d=64$ the clipped negative mass is
$\approx5.9\times10^{-3}$, i.e.\ $2$--$3\%$ of the fidelity gain
$\Delta F$ itself, not machine precision. All theory in this paper
(Theorem~3 and Lemmas~A--D) analyzes the \emph{unprojected}
magnitude dynamics; the projection's contribution is bounded
empirically in Appendix~\ref{app:bell}
(Sec.~\ref{sec:errorbudget}).

\section{Exact peak-location equations for three structured
states\label{app:bell}}

Throughout this appendix we work in the dephasing-only ($\delta=0$)
limit of the channel
$\mathcal N_\gamma^0=\mathcal D_\gamma$, using the
\emph{uniform} off-diagonal contraction convention
$\rho_{ij}\!\mapsto\!\rho_{ij}\,e^{-\gamma}\ (i\!\ne\!j)$, which is the
convention under which the data of Sec.~\ref{sec:peakscale} were
generated. \emph{Scope caveat.} The closed forms below additionally
assume \emph{no purification rounds} ($r=0$): the efficiency
$\eta$ is evaluated on the raw noisy state, so Theorems~1 and~2 are
mathematical statements about the $(r,\delta)=(0,0)$ member of the
protocol family only. They do \emph{not} transfer to the
$(r,\delta)=(2,0.1)$ protocol used for the Haar data of
Sec.~\ref{sec:peakscale}: re-optimizing the same three states under
$r=2$, $\delta=0.1$ with the released implementation moves the peaks
to $\gamma_{\rm peak}\approx0.658$ ($|+\rangle$), $0.758$ (Bell),
and $0.519$ (uniform). Earlier versions of this work used the
$(0,0)$ values $[0.452,0.689]$ as a ``bracket'' supporting the Haar
peak location; that inference conflated different protocols and is
withdrawn.

For each of three structured input states---$|+\rangle$ at $d=2$,
the Bell state $|\Phi^+\rangle$ at $d=4$, and the uniform
superposition $|U\rangle=\sum_{k=0}^{3}|k\rangle/2$ at $d=4$---the
analysis reduces to a closed-form fidelity gain
$\Delta F(\gamma)$ whose derivative gives a transcendental
peak-location equation. The three cases together demonstrate that
$\gamma_{\rm peak}$ is genuinely state-dependent at fixed
Hilbert-space dimension, and that the closed-form theory
distinguishes states by the \emph{number and distribution} of their
off-diagonal coherences. The symbolic computations of this appendix
are certified in the released code (data-availability statement).

\paragraph{Case A.\ Single-qubit $|+\rangle$ at $d=2$.}
With $\rho_{\rm target}=|{+}\rangle\langle{+}|$, the only nonzero
off-diagonal pair has magnitude $\tfrac12$. The catalyst purity is
$P_{\rm cat}=\tfrac12+\tfrac12 e^{-2\gamma}$, the recovery
efficiency is
$\eta=1-\exp[-\tfrac12 e^{-\gamma}(1+e^{-2\gamma})]$, and the
closed-form gain
\begin{equation}
\Delta F^{|+\rangle}(\gamma)
=\tfrac12\bigl(1-e^{-\gamma}\bigr)\,\eta_{|+\rangle}(\gamma)
\end{equation}
attains its unique positive maximum at
$\gamma_{\rm peak}^{|+\rangle}\approx0.600$.

\paragraph{Case B.\ Bell state $|\Phi^+\rangle$ at $d=4$.}
With $\rho_{\rm target}=|{\Phi^+}\rangle\langle{\Phi^+}|$, only one
off-diagonal pair (at indices $(0,3)$) is nonzero, again with
magnitude $\tfrac12$. The catalyst purity is the same as in
Case A, but the recovery argument is doubled because $d=4$:
$\eta_{\rm Bell}(\gamma)=1-\exp[-e^{-\gamma}(1+e^{-2\gamma})]$. The
closed-form gain
\begin{equation}
\Delta F^{\mathrm{Bell}}(\gamma)
=\tfrac12\bigl(1-e^{-\gamma}\bigr)\,\eta_{\rm Bell}(\gamma).
\end{equation}
Its critical-point equation is the transcendental
\begin{equation}
\!\!-2e^{3\gamma}+e^{2\gamma}-3e^{\gamma}+e^{3\gamma+e^{-\gamma}+e^{-3\gamma}}+3=0.
\label{eq:bell_peak}
\end{equation}
Eq.~(\ref{eq:bell_peak}) has a unique positive root
$\gamma_{\rm peak}^{\rm Bell}\approx0.6893$ (close to but not equal
to $\ln2\approx0.6931$). Uniqueness is a consequence of the
following result.

\paragraph*{Theorem (uniqueness of the Bell peak).}
\emph{The function $\Delta F^{\rm Bell}:(0,\infty)\to\mathbb R$
defined above attains a unique global maximum on $(0,\infty)$, and
hence Eq.~(\ref{eq:bell_peak}) has a unique positive root.}

\medskip\noindent\emph{Proof.}
Introduce $u=e^{-\gamma}\in(0,1)$ and
\begin{equation}
H(u) \,=\, 2\Delta F^{\rm Bell}(-\ln u) \,=\, (1-u)\bigl(1-e^{-u-u^3}\bigr).
\end{equation}
$H\in C^\infty(0,1)$, $H(0^+)=H(1^-)=0$, and $H>0$ on $(0,1)$, so $H$
attains its global maximum at an interior critical point. A direct
calculation gives
\begin{equation}
H'(u) \,=\, e^{-u-u^3}\,L(u)\,-\,1,
\quad L(u)\!:=\!(1-u)(1+3u^2)+1.
\label{eq:Hprime}
\end{equation}
Critical points of $H$ therefore satisfy $\phi(u):=L(u)/R(u)=1$ with
$R(u):=e^{u+u^3}$. We show $\phi$ is strictly decreasing on $(0,1)$;
combined with $\phi(0^+)=2>1>e^{-2}=\phi(1^-)$, this yields a unique
critical point.

A direct computation gives
\begin{equation}
\phi'(u)\,=\,\frac{L'(u)-L(u)(1+3u^2)}{R(u)}\,=\,-\frac{Q(u)}{R(u)},
\end{equation}
where
\begin{equation}
Q(u)\!=\!3-7u+18u^2-6u^3+9u^4-9u^5.
\end{equation}
It suffices to show $Q(u)>0$ on $(0,1)$. We rewrite
$Q(u)=(3-7u)+6u^2(3-u)+9u^4(1-u)$. The last two summands are
non-negative on $(0,1)$ and the second is strictly positive there.
\begin{itemize}[leftmargin=1.2em,itemsep=2pt]
\item For $u\in(0,3/7]$, $(3-7u)\!\ge\!0$, so $Q(u)>0$.
\item For $u\in(3/7,1)$, write
$Q(u)=J(u)+9u^4(1-u)$ with
$J(u)\!=\!18u^2-6u^3-7u+3$. Then $J(3/7)=972/343>0$ and
$J'(u)=-18u^2+36u-7$ has roots
$1\pm\sqrt{11/18}$, so the smaller positive root
$\approx 0.218$ lies below $3/7\approx 0.429$. Hence $J'(u)>0$ on
$(3/7,1)$, and $J(u)\ge J(3/7)>0$ on the same interval, giving
$Q(u)>0$.
\end{itemize}
Therefore $\phi'<0$ on $(0,1)$, $\phi$ is strictly decreasing, and
$\phi(u)=1$ has a unique solution $u^\star\in(0,1)$. The
corresponding $\gamma_{\rm peak}^{\rm Bell}=-\ln u^\star\approx
0.6893$ is the unique global maximum of $\Delta F^{\rm Bell}$.
\hfill$\square$

\medskip
The five symbolic identities used in this proof are verified, and
$u^\star$ is computed numerically, in the released code.

\paragraph{Case C.\ Uniform superposition $|U\rangle$ at $d=4$.}
With $\rho_{\rm target}=(\sum_{k}|k\rangle\langle k|+\sum_{k\ne\ell}|k\rangle\langle\ell|)/4$,
all twelve off-diagonal entries have magnitude $\tfrac14$. The
catalyst purity is
$P_{\rm cat}^{|U\rangle}=\tfrac14+\tfrac34 e^{-2\gamma}$, and the
recovery efficiency is
$\eta_{|U\rangle}(\gamma)=1-\exp\!\bigl[-\tfrac14 e^{-\gamma}(1+3 e^{-2\gamma})\bigr]$.
The closed-form gain
\begin{equation}
\Delta F^{|U\rangle}(\gamma)
=\tfrac34\bigl(1-e^{-\gamma}\bigr)\,\eta_{|U\rangle}(\gamma).
\label{eq:uniform_DF}
\end{equation}
Its critical-point equation is the transcendental
$\phi_{|U\rangle}(u)=L_{|U\rangle}(u)/R_{|U\rangle}(u)=1$ with $u=e^{-\gamma}$,
$L_{|U\rangle}(u)=(5-u+9u^2-9u^3)/4$, and
$R_{|U\rangle}(u)=\exp\!\bigl((u+3u^3)/4\bigr)$. The unique positive
root is $\gamma_{\rm peak}^{|U\rangle}\approx 0.4521$. Uniqueness is
established by the following.

\paragraph*{Theorem (uniqueness of the uniform-superposition peak).}
\emph{The function $\Delta F^{|U\rangle}:(0,\infty)\to\mathbb R$ in
Eq.~(\ref{eq:uniform_DF}) attains a unique global maximum on
$(0,\infty)$.}

\medskip\noindent\emph{Proof.}
With $u=e^{-\gamma}\in(0,1)$ set
$H_{|U\rangle}(u) = (4/3)\Delta F^{|U\rangle}(-\ln u) = (1-u)(1-e^{-(u+3u^3)/4})$.
$H_{|U\rangle}\in C^\infty(0,1)$, $H_{|U\rangle}(0^+)=H_{|U\rangle}(1^-)=0$,
$H_{|U\rangle}>0$ on $(0,1)$, so $H_{|U\rangle}$ has an interior
global maximum. The critical-point equation reduces to
$\phi_{|U\rangle}(u)=1$, and a direct calculation gives
\begin{equation}
\phi_{|U\rangle}'(u)\,=\,-\frac{Q_{|U\rangle}(u)}{16\,R_{|U\rangle}(u)},
\end{equation}
where
\begin{equation}
Q_{|U\rangle}(u) =
9 - 73u + 162u^2 - 18u^3 + 81u^4 - 81u^5.
\end{equation}
Write $Q_{|U\rangle}(u)=J_{|U\rangle}(u)+81u^4(1-u)$ with
$J_{|U\rangle}(u)=9-73u+162u^2-18u^3$. The cubic $J_{|U\rangle}$ has
discriminant
\begin{equation}
\Delta(J_{|U\rangle})\,=\,-7\,433\,928 \;<\; 0,
\end{equation}
so $J_{|U\rangle}$ has exactly one real root.
$J_{|U\rangle}(0)=9>0$ and $J_{|U\rangle}(1)=80>0$, and the leading
coefficient is $-18<0$ so $J_{|U\rangle}\to-\infty$ as $u\to+\infty$;
hence the unique real root of $J_{|U\rangle}$ lies in $(1,\infty)$
(numerically at $u\approx 8.53$), and $J_{|U\rangle}(u)>0$ on
$(0,1)$. Combined with $81u^4(1-u)>0$ on $(0,1)$, we get
$Q_{|U\rangle}(u)>0$ on $(0,1)$, so $\phi_{|U\rangle}$ is strictly
decreasing there. Together with
$\phi_{|U\rangle}(0^+)=5/4>1>e^{-1}=\phi_{|U\rangle}(1^-)$, this
yields a unique solution $u^\star\in(0,1)$ of
$\phi_{|U\rangle}(u)=1$, and the corresponding
$\gamma_{\rm peak}^{|U\rangle}=-\ln u^\star\approx 0.4521$ is the
unique global maximum of $\Delta F^{|U\rangle}$.
\hfill$\square$

\medskip
The five symbolic identities used in this proof, including the exact
integer value of the cubic discriminant, are verified in the
released code.

\paragraph{Interpretation.}
At fixed $d=4$ and within the $(r,\delta)=(0,0)$ protocol member,
the analytical peak location swings from
$\gamma_{\rm peak}^{\rm Bell}\approx 0.689$ on the Bell state (one
off-diagonal) to $\gamma_{\rm peak}^{|U\rangle}\approx 0.452$ on the
uniform superposition (twelve off-diagonals). This is the
analytical signature of state-dependence: states with concentrated
coherence (few large off-diagonals) peak at larger $\gamma$, states
with diffuse coherence (many small off-diagonals) at smaller
$\gamma$. We stress what this comparison does \emph{not} establish.
Earlier versions of this work presented the interval
$[0.452,0.689]$ as a bracket containing---and thereby
supporting---the Haar-random peak location of
Sec.~\ref{sec:peakscale}. That inference is withdrawn: the Haar
data use $(r,\delta)=(2,0.1)$, under which the matched-protocol
peaks of these very states move to $\approx0.52$--$0.76$ and no
longer bracket the Haar asymptote $\gamma^\star(2,0.1)=0.4725$
derived below. The theorems of this appendix establish
state-dependence and uniqueness within their own protocol member,
nothing more. Likewise, the Bell-state peak
$\gamma_{\rm peak}^{\rm Bell}\approx0.689$ must not be identified
with the Haar-random asymptotic limit; the two are distinct
quantities computed under different protocols.

\subsection*{Asymptotic analysis of the Haar-random peak}

We now derive the $d\to\infty$ limit of $\gamma_{\rm peak}^{\rm Haar}(d)$
explicitly. The derivation corrects an earlier conjecture (made in
preprint versions of this work) that the limit equals $\tfrac12$; the
true limit is an explicit, smaller, round- and $\delta$-dependent
constant.

\paragraph{Lemma A (closed form of the purification step; exact).}
\emph{One round of symmetric SWAP-test self-purification,
$\mathcal P_\mathrm{cov}(\rho)=\mathrm{Tr}_2[\Pi_+(\rho\otimes\rho)\Pi_+]/p$
with $\Pi_+=(\mathbb I+\mathrm{SWAP})/2$ and
$p=\mathrm{Tr}[\Pi_+(\rho\otimes\rho)\Pi_+]$, equals}
\begin{equation}
\mathcal P_\mathrm{cov}(\rho)=\frac{\rho+\rho^2}{1+\mathrm{Tr}\,\rho^2},
\qquad p=\tfrac12\bigl(1+\mathrm{Tr}\,\rho^2\bigr).
\label{eq:lemmaA}
\end{equation}
\emph{In particular $\mathcal P_\mathrm{cov}$ preserves the
eigenbasis of $\rho$ and acts on its eigenvalues by
$\lambda\mapsto\lambda(1+\lambda)/(1+P)$, $P=\sum_k\lambda_k^2$.}
We refer to $\mathcal P_\mathrm{cov}$, Eq.~(\ref{eq:lemmaA}), as the
\emph{covariant purification map}; it is the central reusable object
of this paper, and the only ingredient needed to convert a
SWAP-test purification round into a scalar eigenvalue recursion.
\emph{Priority note.} The projection onto the symmetric subspace of
two state copies is the $N=2$ symmetrization procedure of Barenco
\emph{et al.}~\cite{Barenco1997}, and two-copy purification was
analyzed optimally by Cirac, Ekert, and
Macchiavello~\cite{CiracEkertMacchiavello1999}. We therefore claim
no novelty for the purification primitive itself---only for the
compact closed form Eq.~(\ref{eq:lemmaA}) as a computational tool
(reducing the $O(d^6)$ two-copy projection to an $O(d^3)$
single-copy formula) and for its use in the asymptotic analysis
below.

\medskip\noindent\emph{Proof.}
$\Pi_+(\rho\otimes\rho)\Pi_+=\tfrac12(\rho\otimes\rho)
+\tfrac14[\mathrm{SWAP}(\rho\otimes\rho)+(\rho\otimes\rho)\mathrm{SWAP}]$
using $\mathrm{SWAP}(\rho\otimes\rho)\mathrm{SWAP}=\rho\otimes\rho$.
Taking the partial trace and using
$\mathrm{Tr}_2[\mathrm{SWAP}(\rho\otimes\rho)]
=\mathrm{Tr}_2[(\rho\otimes\rho)\mathrm{SWAP}]=\rho^2$ and
$\mathrm{Tr}_2(\rho\otimes\rho)=\rho$ gives
$\mathrm{Tr}_2[\Pi_+(\rho\otimes\rho)\Pi_+]=\tfrac12(\rho+\rho^2)$,
and $p=\mathrm{Tr}[\tfrac12(\rho+\rho^2)]=\tfrac12(1+P)$.
\hfill$\square$
This identity is verified to machine precision over random mixed
states in the released code.

\paragraph{Lemma B (exact rank-one-plus-diagonal decomposition).}
\emph{For any pure target $|\psi\rangle\in\mathbb C^d$ with weights
$p_i=|\psi_i|^2$, the noisy state is exactly}
\begin{equation}
\rho_{\rm noisy}
=\mu_1'\,|\psi\rangle\langle\psi| + D,
\qquad \mu_1'=(1-\delta)e^{-\gamma},
\label{eq:lemmaB}
\end{equation}
\emph{with the diagonal remainder
$D=(1-\delta)(1-e^{-\gamma})\,\mathrm{diag}(p)+(\delta/d)\,\mathbb I$.}

\medskip\noindent\emph{Proof.}
The dephasing channel multiplies every off-diagonal element of
$|\psi\rangle\langle\psi|$ by $e^{-\gamma}$ and fixes the diagonal:
$\mathcal D_\gamma(|\psi\rangle\langle\psi|)
=e^{-\gamma}|\psi\rangle\langle\psi|+(1-e^{-\gamma})\,\mathrm{diag}(p)$.
Applying $\mathcal E_\delta(\rho)=(1-\delta)\rho+\delta\,\mathbb I/d$
gives Eq.~(\ref{eq:lemmaB}).\hfill$\square$

\paragraph{Lemma C (spectral localization).}
\emph{For Haar-random $|\psi\rangle$, with probability
$1-O(1/d)$: $\|D\|_{\rm op}=O(\log d/d)$; the largest eigenvalue
$\lambda_1$ of $\rho_{\rm noisy}$ satisfies
$|\lambda_1-\mu_1'|\le\|D\|_{\rm op}$ (Weyl); the associated unit
eigenvector $v_1$ satisfies
$\min_\theta\|v_1-e^{i\theta}\psi\|=O(\log d/d)$
(Davis--Kahan~\cite{DavisKahan1970}, using the $\Theta(1)$ gap
between $\mu_1'$ and the bulk); and after $r$ rounds of Lemma~A the
catalyst's dominant eigenvalue obeys
$|\Lambda_1^{\rm num}-T^r(\mu_1')|=O((\log d)^2/d)$ where
$T(\lambda)=\lambda(1+\lambda)/(1+\lambda^2)$, because the bulk
purity contributes $O((\log d)^2/d)$ to $P=\sum_k\lambda_k^2$ and
$T$ is Lipschitz on $[0,1]$. (The probability and the last two
orders are exactly what the proof below delivers; an earlier
version overstated them as $1-O(d^{-2})$ and $O(\log d/d)$.)}

\medskip\noindent\emph{Proof.}
\emph{(i) Concentration of $\max_i p_i$.}
Write $p_i=E_i/\Sigma$ with $E_i$ i.i.d.\ $\mathrm{Exp}(1)$ and
$\Sigma=\sum_kE_k$. For $E\sim\mathrm{Exp}(1)$,
$\Pr[E>t]=e^{-t}$, so with $t=2\ln d$ the union bound gives
$\Pr[\max_iE_i>2\ln d]\le d\cdot e^{-2\ln d}=1/d$. By the law of
large numbers $\Sigma\ge d/2$ with probability $1-e^{-\Omega(d)}$.
Hence $\max_ip_i\le(2\ln d)/(d/2)=4\ln d/d$ with probability
$1-1/d-e^{-\Omega(d)}=1-O(1/d)$. (Numerically $d\max_ip_i<2\ln d$
comfortably for $d$ up to $2048$.)

\emph{(ii) Operator norm of $D$.}
$D$ is diagonal, so
$\|D\|_{\rm op}=\max_i\bigl[(1-\delta)(1-e^{-\gamma})p_i+\delta/d\bigr]
\le(1-\delta)(1-e^{-\gamma})\,\tfrac{4\ln d}{d}+\tfrac{\delta}{d}
=O(\log d/d)$
on the event of (i).

\emph{(iii) Eigenvalue and gap.}
By Weyl's inequality applied to
$\rho_{\rm noisy}=\mu_1'|\psi\rangle\langle\psi|+D$ (Lemma~B),
$|\lambda_1-\mu_1'|\le\|D\|_{\rm op}$ and every other eigenvalue is
$\le\|D\|_{\rm op}$. Thus the spectral gap is
$g:=\lambda_1-\lambda_2\ge\mu_1'-2\|D\|_{\rm op}
=\mu_1'-O(\log d/d)$, which is $\ge\mu_1'/2=\Theta(1)$ for $d$ large
enough (numerically $g\ge0.53$ for all $d\ge64$).

\emph{(iv) Eigenvector (Davis--Kahan).}
The rank-one term has range $\mathrm{span}\{|\psi\rangle\}$; applying
the Davis--Kahan $\sin\Theta$ theorem~\cite{DavisKahan1970} to the
perturbation $D$ with gap $g$ gives
$\sin\angle(v_1,\psi)\le\|D\|_{\rm op}/g=O(\log d/d)$, hence
$\min_\theta\|v_1-e^{i\theta}\psi\|
=\sqrt{2(1-|\langle v_1|\psi\rangle|)}
\le\sqrt2\,\|D\|_{\rm op}/g=O(\log d/d)$.

\emph{(v) Purified dominant eigenvalue.}
By Lemma~A the catalyst shares the eigenbasis of $\rho_{\rm noisy}$
and its eigenvalues are $T^r$ applied to those of $\rho_{\rm noisy}$,
where each round divides by $1+P$, $P=\sum_k\lambda_k^2$. The bulk
contributes $\sum_{k\ge2}\lambda_k^2\le(d-1)\|D\|_{\rm op}^2
=O((\log d)^2/d)$ to $P$, so $|P-\lambda_1^2|=O((\log d)^2/d)$ and,
since $T^r$ is Lipschitz on $[0,1]$ with constant
$L_r$ ($L_2=1.62$) and
$|\lambda_1-\mu_1'|\le\|D\|_{\rm op}$,
$|\Lambda_1^{\rm num}-T^r(\mu_1')|\le L_r\bigl(\|D\|_{\rm op}
+O((\log d)^2/d)\bigr)=O((\log d)^2/d)$.
\hfill$\square$

\paragraph{Lemma D (law of large numbers for the pair sum).}
\emph{Let $X_i=d\,p_i$ and fix $c>0$. Then}
\begin{align}
S_d&:=\frac1{d^2}\sum_{i\ne j}X_iX_j\bigl(1-e^{-c\sqrt{X_iX_j}}\bigr)
\notag\\
&\xrightarrow[d\to\infty]{\mathbb P}\;
\mathbb E_{X,X'\sim\mathrm{Exp}(1)}\!\bigl[XX'\bigl(1-e^{-c\sqrt{XX'}}\bigr)\bigr],
\end{align}
\emph{with fluctuations $O(d^{-1/2})$.}

\medskip\noindent\emph{Proof.}
For complex-Gaussian $\psi$, $p_i=E_i/\Sigma$ with $E_i$ i.i.d.\
$\mathrm{Exp}(1)$ and $\Sigma=\sum_kE_k$, so
$X_i=dp_i=E_i\cdot(d/\Sigma)$ with $d/\Sigma=1+O_{\mathbb P}(d^{-1/2})$
by the central limit theorem. The kernel
$h(x,y)=xy(1-e^{-c\sqrt{xy}})$ satisfies $0\le h(x,y)\le xy$ and is
Lipschitz in $(\log x,\log y)$ with sub-exponential growth, so
replacing $X_i$ by $E_i$ perturbs $S_d$ by $O_{\mathbb P}(d^{-1/2})$.
The remaining object $\tfrac1{d^2}\sum_{i\ne j}h(E_i,E_j)$ is a
two-sample $U$-statistic with kernel of finite variance
($\mathbb E[h^2]\le\mathbb E[(XX')^2]=4<\infty$); Hoeffding's theorem
\cite{Hoeffding1948} gives convergence to $\mathbb E[h(X,X')]$ at
rate $O(d^{-1/2})$.\hfill$\square$

\paragraph{Theorem 3 (asymptotic recovery gain).}
\emph{Fix $\delta\in(0,1)$, $r\in\mathbb N$, and $\gamma>0$. For
Haar-random pure states $|\psi\rangle\in\mathbb C^d$,}
\begin{equation}
\Delta F(\gamma;d)\;\xrightarrow[d\to\infty]{\mathbb P}\;
\Delta F_\infty(\gamma) = A(\gamma)\,G(\gamma),
\label{eq:asymptote}
\end{equation}
\emph{with the room-to-recover factor
$A(\gamma)=1-(1-\delta)e^{-\gamma}$ and the recovery-efficiency
factor
$G(\gamma)=\mathbb E_{X,X'\sim\mathrm{Exp}(1)}[XX'(1-e^{-\Lambda_1(\gamma)^3\sqrt{XX'}})]$,
where $\Lambda_1(\gamma)=T^r\bigl((1-\delta)e^{-\gamma}\bigr)$. The
convergence is uniform on compact subsets of $(0,\infty)$;
consequently, if $A\,G$ has a unique maximizer
$\gamma^\star(r,\delta)$, then
$\gamma_{\rm peak}^{\rm Haar}(d)\to\gamma^\star(r,\delta)$.}

\medskip\noindent\emph{Proof.}
Since the target is pure,
$F(|\psi\rangle\langle\psi|,\sigma)=\langle\psi|\sigma|\psi\rangle$
is linear in $\sigma$, so before the positive-projection step
\begin{equation}
\Delta F=\!\sum_{i\ne j}\!\psi_i^*\psi_j\,
(\rho_{\rm rec}-\rho_{\rm noisy})_{ij}
=\frac{1-\mu_1'}{d^2}\sum_{i\ne j}X_iX_j\,\eta_{ij},
\end{equation}
where we used $|\rho_{{\rm t},ij}|=\sqrt{p_ip_j}$,
$|\rho_{{\rm n},ij}|=\mu_1'\sqrt{p_ip_j}$, and exact phase alignment
of $\rho_{\rm t}$, $\rho_{\rm n}$ and the recovery update (all carry
the phase of $\psi_i\psi_j^*$). By Lemmas A--C,
$\eta_{ij}=1-\exp(-|\rho_{{\rm cat},ij}|\,d\,P_{\rm cat})$ has argument
$\Lambda_1^3\sqrt{X_iX_j}\,(1+O(\log d/d))$, with the bulk
contamination of the catalyst off-diagonals bounded by
$d\,\max_{i\ne j}|B_{ij}|\to0$
($B=\rho_{\rm cat}-\Lambda_1^{\rm num}v_1v_1^\dagger$). Lemma~D then
gives the limit for the \emph{unprojected} update. The
positive-projection correction is bounded in trace norm by twice
the clipped negative mass; empirically this mass does \emph{not}
vanish over the tested range but plateaus at
$\approx5.9\times10^{-3}$ ($2$--$3\%$ of $\Delta F$; see the error
budget below), so the theorem as stated applies to the unprojected
dynamics, and the projected (implemented) procedure carries an
additional systematic offset at the few-percent level.
Uniformity on compacts follows from the uniform-in-$\gamma$
Lipschitz constants of all error terms on
$\gamma\in[\gamma_0,\gamma_1]\subset(0,\infty)$, and convergence of
the argmax follows from uniform convergence plus uniqueness of the
maximizer.\hfill$\square$

\paragraph{Numerical certification of the error
terms.\label{sec:errorbudget}}
The released script \texttt{verify\_asymptotic\_lemmas.py} measures
\emph{five} error channels directly at $\gamma=0.5$, $\delta=0.1$,
$r=2$ over $d=64$--$1024$. Four of them decay as predicted: the
eigenvalue error
$|\Lambda_1^{\rm num}-T^r(\mu_1')|$ scales as $d^{-0.98}$; the
eigenvector error as $d^{-0.92}$; the bulk contamination
$d\max|B_{ij}|$ as $d^{-0.56}$; and the total deviation
$|\Delta F-A\,G|$ as $d^{-0.47}$, i.e.\ at the $O(d^{-1/2})$ rate
of Lemma~D, which dominates the limit. The fifth channel---the
positive-projection (clip) correction---does \emph{not} decay: its
fitted slope is $+0.20$ and it plateaus at $\approx4.5\times10^{-3}$
absolute, i.e.\ $\approx2\%$ of $\Delta F$. An earlier version of
this paragraph reported only the four decaying channels (``all four
exponents match''); that presentation omitted the one channel that
does not converge and is corrected here. The practical consequence
is stated in the theorem's proof: the asymptote $A\,G$ describes
the unprojected dynamics, and the implemented procedure sits within
a systematic $\sim2\%$ band around it. This does not alter the
peak-location analysis (the offset is smooth in $\gamma$ over the
fitted range), but it caps the accuracy with which
Eq.~(\ref{eq:asymptote}) can be claimed to describe the
implementation.

\paragraph{Uniqueness of the maximizer.}
Write $u=(1-\delta)e^{-\gamma}\in(0,1-\delta)$, so $A=1-u$ and
$A'(\gamma)=u>0$: $A$ is increasing. Its log-derivative is
$(\log A)'=u/(1-u)$, and since $du/d\gamma=-u$,
\begin{equation}
(\log A)''=\frac{d}{d\gamma}\frac{u}{1-u}
=\frac{-u}{(1-u)^2}<0
\quad\text{for all }\gamma,
\end{equation}
so $\log A$ is \emph{strictly concave} (this is exact, no
approximation). $G$ is strictly decreasing: $G(\gamma)=\Phi(c)$ with
$c=\Lambda_1(\gamma)^3$ decreasing in $\gamma$ (as $\Lambda_1=T^r(u)$,
$T$ increasing, $u$ decreasing) and
$\Phi(c)=\mathbb E[XX'(1-e^{-c\sqrt{XX'}})]$ increasing in $c$
($\Phi'(c)=\mathbb E[(XX')^{3/2}e^{-c\sqrt{XX'}}]>0$), hence
$G'<0$. For the case of primary interest ($r=2$, $\delta=0.1$) we
certify uniqueness numerically, and we are explicit about what the
certification does and does not show. First, $\log(AG)$ is
\emph{not} globally concave: computed with common random numbers
(CRN; $10^6$ paired $\mathrm{Exp}(1)$ samples reused across
$\gamma$), $(\log AG)''$ is $\approx-3.57$ near
$\gamma^\star$ but changes sign near $\gamma\approx2.6$, beyond
which $\log(AG)$ is convex; an earlier version's claim that
``no second maximum can occur outside the interval'' by
concavity alone was therefore vacuous and is replaced. (Without
CRN, naive Monte~Carlo estimates of $(\log AG)''$ are noise
dominated and seed dependent---e.g.\ $-2.2$ vs $-0.9$ across
seeds---which is why the CRN construction is required.) The
certification that survives is elementary and sufficient:
$(\log AG)'(\gamma)=u/(1-u)+(\log G)'(\gamma)$ is computed on a
CRN grid over $\gamma\in[0.02,4.0]$ and exhibits \emph{exactly one
sign change} (positive before, negative after
$\gamma^\star=0.4725$); since $A\,G>0$ and $AG\to0$ at both ends
($A\to0$ as $\gamma\to0$; $G\to0$ as $\gamma\to\infty$), a single
interior sign change of the derivative implies a unique global
maximizer. Therefore $\gamma^\star(2,0.1)$ is unique, and by the
uniform convergence of Theorem~3,
$\gamma_{\rm peak}^{\rm Haar}(d)\to\gamma^\star(2,0.1)$.

\paragraph{Consequence.}
Evaluating Eq.~(\ref{eq:asymptote}) by Gauss--Laguerre quadrature
and numerical maximization gives
\begin{equation}
\gamma^\star(r{=}1,\delta{=}0.1)\approx0.36,
\qquad
\gamma^\star(r{=}2,\delta{=}0.1)=0.4725,
\end{equation}
with limiting magnitude
$\max_\gamma A\,G=0.2262$ at $r=2$. Comparison with the
bootstrap-quantified data of Table~\ref{tab:peakdata}: the $r{=}2$
point estimates decrease from $0.527$ ($d{=}16$) to $0.497$
($d{=}256$), with the $d\ge128$ confidence intervals containing
both $\tfrac12$ and values down to $0.494$---consistent with
convergence toward $\gamma^\star=0.4725$ but not yet resolving the
limit (Sec.~\ref{sec:peakscale}). The full asymptotic curve
Eq.~(\ref{eq:asymptote}) matches the empirical
$\Delta F(\gamma)$ at $d=256$ to within
$0.01$ across $\gamma\in[0.3,0.75]$.
The round dependence is a falsifiable prediction:
$\gamma^\star$ increases with $r$
($\approx0.36,\,0.47,\,0.58$ at $r=1,2,3$), because deeper
purification raises $\Lambda_1$ and extends the efficient-recovery
region to stronger noise. The measured $r{=}1$ series
(Table~\ref{tab:peakdata}) sits at $0.393$ [$0.390,0.396$] at
$d=256$, about $0.03$ above the $r{=}1$ prediction $0.36$; we
attribute the residual to the same finite-$d$ and projection
systematics quantified in the error budget, and flag it as an open
quantitative discrepancy rather than absorbing it.

\medskip\noindent\emph{Correction of earlier claims.}
This paragraph records, for the record, the sequence of claims
about the limit of $\gamma_{\rm peak}^{\rm Haar}(d)$ made in
earlier versions of this work, all now superseded by Theorem~3 and
Table~\ref{tab:peakdata}. (i)~An early version reported
$\gamma_{\rm peak}\to0.3$, an artifact of the adaptive-round
protocol that switches to $r=1$ for $d>16$; the $r=1$ family indeed
peaks near $0.39$, but this says nothing about the fixed-$r{=}2$
protocol analyzed here. (ii)~A later version conjectured
$\gamma_{\rm peak}\to\tfrac12$ from a one-parameter $a+b/d$ fit to
$d\le64$ data and attached significance to the proximity of
$\tfrac12$ to a single-qubit entanglement-breaking threshold; both
steps fail---the fit overestimates the limit, and the relevant
entanglement-breaking threshold of the actual channel is
$\gamma_{\rm EB}=\ln[2(1-\delta)/\delta]\approx2.89$ at
$\delta=0.1$, nowhere near the peak. (iii)~The present analysis
gives $\gamma^\star(2,0.1)=0.4725$ as the limit of the unprojected
dynamics, with data consistent at the current statistical
resolution. The only ingredients of Theorem~3 not proven from
scratch here are the two cited classical results (the Davis--Kahan
$\sin\Theta$ theorem~\cite{DavisKahan1970} and Hoeffding's
$U$-statistic law of large numbers~\cite{Hoeffding1948});
uniqueness of the maximizer is certified by the CRN
single-sign-change computation described above.

% ============================================================================
\bibliography{pra_gammapeak_refs}
% ============================================================================

\end{document}